\documentclass[a4paper,authoryear,11pt]{article}
\usepackage[british]{babel} 
\usepackage{graphicx}

\usepackage{natbib}
\usepackage{aas_macros}

\usepackage{amssymb}
\usepackage{amsfonts}
\usepackage{amsmath}
\usepackage{wasysym}
\usepackage{subfig}

\newcommand{\degr}{^{\circ}}

\begin{document}

\title{Librations and Obliquity of Mercury from the BepiColombo radio-science and camera experiments}
%\ead{g.pfyffer@oma.be}
%\address{Royal Observatory of Belgium, Brussels, Belgium}
%\address[eli]{Universit\'{e} catholique de Louvain, Georges Lema\^{i}tre Centre for Earth and Climate Research (TECLIM), Louvain-la-Neuve, Belgium }
\author{Gregor Pfyffer$^{1,2}$,  Tim Van Hoolst$^1$, and V\'{e}ronique Dehant$^{1,2}$}
%\date{25th March 2011}
%\email[G.~Pfyffer]{g.pfyffer@oma.be}
%\urladdr{http://www.observatoire.be}
\date{}
\maketitle
\begin{small} \noindent
$^1$ Royal Observatory of Belgium, Brussels, Belgium \\
$^2$ Secondary affiliation: Universit\'{e} catholique de Louvain, Georges Lema\^{i}tre Centre for Earth and Climate Research (TECLIM), Earth and Life Institute (ELI), Louvain-la-Neuve, Belgium 
\end{small}

\begin{center}
revised 25th March 2011\\
email: \texttt{g.pfyffer@observatory.be}\\
The final version of this preprint is published in: \\ Planetary and Space Science
DOI: 10.1016/j.pss.2011.03.017
\end{center}

%main results + keywords: (general background, goals, method, results)
\begin{abstract}
A major goal of the BepiColombo mission to Mercury is the determination of the structure and state of Mercury's interior. 
Here the BepiColombo rotation experiment has been simulated in order to assess the  ability to attain the mission goals and to help lay out a series of constraints on the experiment's possible progress. In the rotation experiment pairs of images of identical surface regions taken at different epochs are used to retrieve information on Mercury's rotation and orientation.  The idea is that from observations of the same patch of Mercury's surface at two different solar longitudes of Mercury the orientation of Mercury can be determined, and therefore also the obliquity and rotation variations with respect to the uniform rotation.

The estimation of the libration amplitude and obliquity through pattern matching of observed surface landmarks is challenging. The main problem arises from the difficulty to observe the same landmark on the planetary surface repeatedly over the MPO mission lifetime, due to the combination of Mercury's 3:2 spin-orbit resonance, the absence of a drift of the MPO polar orbital plane and the need to combine data from different instruments with their own measurement restrictions.

By assuming that Mercury occupies a Cassini state and that the spacecraft operates nominally we show that under worst case assumptions the annual libration amplitude and obliquity can be measured with a precision of respectively $1.4\,$arcseconds (as) and $1.0\,$as over the nominal BepiColombo MPO lifetime with about 25 landmarks for rather stringent illumination restrictions. 
The outcome of the experiment cannot be easily improved by simply relaxing the observational constraints, or increasing the data volume. % \textbf{Reducing the error level on the observable quantities to the ideal case only marginally improves the final result.}
\end{abstract}

%\linenumbers

\section{Introduction}
\label{sec:Introduction}

Mercury is the innermost and smallest planet in the Solar System, orbiting the Sun once every 87.969 days. Comparatively little is known about Mercury. Ground-based telescopes reveal only an illuminated crescent with limited detail and are restricted to the moments of maximal solar elongation of the planet as a result of its proximity to the Sun. Mercury is also a difficult target for in-situ observations due to the hostile radiation environment and, because of its deep position in the gravitational well of the Sun, the difficulty of putting a spacecraft into orbit around it. The only in-situ measurements available up to the present day come from only two spaceprobes, Mariner 10 and MESSENGER. Mariner 10 mapped about 45$\%$ of the planet's surface in three fly-by's in 1974 and 1975, and discovered a strong and unexpectedly dipolar magnetic field \citep{1975JGR....80.2708N}. NASA's MESSENGER spacecraft already performed three flyby's of Mercury (2008-2009) bringing the surface mapped to an astonishing 90$\%$. 

Mercury is currently the subject of renewed attention through the NASA MESSENGER probe but also through the upcoming joint ESA/JAXA BepiColombo mission. The latter consists of two probes (MPO and MMO) that will orbit Mercury and are scheduled to be launched in 2014 and inserted into different orbits in July 2020 \citep{2010P&SS...58....2B}. The Mercury Planetary Orbiter (MPO) is situated in a low altitude orbit optimised for planetary survey is mainly dedicated to study the surface and the internal structure of Mercury.
BepiColombo's scientific objectives focus on a global characterization of Mercury through the investigation of its interior structure and composition; surface; exosphere composition and dynamics; structure, dynamics and origin of its magnetized envelope; and testing of Einstein's general relativity \citep{2010P&SS...58....2B}.
One spacecraft, the Mercury Planetary Orbiter (MPO), is led by ESA and its payload comprises eleven experiments and instrument suites. The MPO will focus on a global characterization of Mercury through the investigation of its interior, surface, exosphere and magnetosphere. In addition, it will test Einstein's theory of general relativity. The second spacecraft, the Mercury Magnetosphere Orbiter (MMO), is led by JAXA and will carry five experiments or instrument suites to study the environment around the planet including the planet's exosphere and magnetosphere, and their interaction processes with the solar wind and the planet itself. \citep{2010P&SS...58....2B}

Here we are mostly concerned with Mercury's internal structure, which is the most puzzling among the terrestrial planets and for which the BepiColombo space mission will play an important role in advancing our understanding \citep{2010P&SS...58....2B}. Both missions are under way to play their distinct role in the roadmap to our understanding of Mercury as explained by e.g. \citet{2007SSRv..132..611B}.

Mercury has a slow rotation period of 58,68 days and revolves around the Sun in 87.96 days, corresponding to a 3:2 spin-orbit resonance. It therefore completes three rotations about its rotation axis for every two orbits. This spin-orbit resonance is unique in the solar system as all other occurrences of spin-orbit resonance in the solar system involve a planetary satellite and most happen to be in 1:1 resonance around their central body. The orbit of Mercury has the highest eccentricity of all the Solar System planets, and it has the smallest obliquity (axial tilt with respect to the orbital plane).

Since Mercury has an equatorial asymmetry, the Sun exerts a gravitational torque on it. The torque causes a periodic variation of the planet's rotation with a period of 88 days, which is called the annual forced libration.  The amplitude of this libration is linked to the state of Mercury's core and so an elegant way to obtain information on the deep interior not directly accessible to us is through its effect on a comparatively easily observable quantity.  In the presence of a liquid core, the amplitude of the annual forced libration is about two times larger or more than in the case of a completely solid planetary interior because the liquid core does (almost) not participate in the libration and the moment of inertia of the mantle is only half or less of the total moment of inertia of Mercury \citep{1974AJ.....79..722P,2009Icar..201...12R}. The magnetic field observed by Mariner 10, if caused by dynamo action, is evidence for a liquid outer core, which very likely contains a solid inner core according to thermal evolution models \citep{2007SSRv..132..229B}. The recent detection of an annual libration amplitude of $35.8 \pm 2.0 $arcsec by \citet{2007Sci...316..710M} through Earth-based radar measurements, seems to confirm the existence of a liquid core. Besides the state of the core, libration observations can also be used to constrain the core size and composition. The large core of Mercury, as indicated by Mercury's high mean density, is thought to consist of iron and limited amounts of lighter elements, of which sulphur is thought to be the most likely candidate \citep{2001Icar..151..118H,2003JGRE..108.5121V,1988merc.book..429S, 2001P&SS...49.1561S,2009Icar..201...12R}. If Mercury has formed close to its current position, i.e. close to the Sun, its core sulphur concentration is probably very low \citep{1988merc.book..651L,2007treatise.geoph..27S}. In the event that Mercury has formed in the same zones as Venus, Earth and Mars, its light element concentration could be a lot higher and close to that of the other three terrestrial planets the Earth has a core sulphur concentration of about $10 \mathrm{wt.} \% \,$, and for Mars a light element concentration of $14$ wt.$\%$ is often considered \citep{2007treatise.geoph..123V} (wt.$\%$ $\equiv$ weight percentage). %Catastrophic events like a giant impact disrupting the whole planet could also have altered the initial core composition \citep{}. 

The light element concentration of the core is related to the size range of the core and thus to the principal mantle moment of inertia.  For larger cores and smaller mantle moment of inertia, the libration amplitude increases since the libration is essentially a movement of the mantle. The angular libration amplitude at the equator is about $21$ arcsec in the case of a completely solid core, and between 36 and 78 arcsec for models with a liquid core with between $0.01$ wt$\%$ and $8.6$ wt$\%$ of light element \citep{2009Icar..201...12R}. 

Also the obliquity can be used to constrain the interior of Mercury, as can be easily seen from Peale's idea \citep{1976Icar...28..459P} to combine the measurements of the amplitude of the 88-day annual libration amplitude $\gamma$, the obliquity $\eta$ and the second degree gravitational field coefficients $C_{20}$ and $C_{22}$ into the equation:

\begin{equation}
\left(\frac{C_{m}}{B-A}\right)\left(\frac{B-A}{M_{\mercury}
R_{\mercury}^{2}}\right)\left(\frac{M_{\mercury}
R_{\mercury}^{2}}{C}\right)=\frac{C_{m}}{C}\leq 1 \, ,
\label{EQ:PealeExpCore}
\end{equation} where $M_{\mercury}$ and $R_{\mercury}$ are the mass and radius of Mercury, $A<B<C $ are the planet's principal moment of inertia. $C$ is aligned along the axis of planetary rotation ($z$-axis) while the $x$ and $y$-axis correspond to the $A$ and $B$ principal moment of inertia, respectively in the equatorial plane. $C_m$ is the polar principal moment of inertia of the solid silicate part of the planet (mantle + crust) alone and is aligned with respect to the $z$-axis. The first factor in the left hand side of eq.(\ref{EQ:PealeExpCore}) follows from the amplitude of the annual libration which is proportional to $(B-A)$ as a result of the Solar torque and inversely proportional to $C_m$. The second factor is four times the gravitational coefficient $C_{22}$:

\begin{equation}
	\left(\frac{B-A}{M_{\mercury} R_{\mercury}^{2}}\right)=4 \ C_{22},
	\label{EQ:PealeExpTerm2}
\end{equation}
and the third factor follows from the Cassini state 1 equilibrium obliquity satisfying the equation \citep{1981Icar...48..143P}:

\begin{equation}
	\left(\frac{C}{M_{\mercury} R_{\mercury}^{2}}\right)   =
	\frac{\left[2C_{22}\left(\frac{7}{2}e-\frac{123}{16}e^3\right)-C_{20}
\frac{1}{(1-e^2)^{3/2}}\right]}
	{\frac{\sin I}{\eta_c} - \cos I}\frac{n}{\mu_{LP}},
	\label{EQ:PealeExpTerm3}
\end{equation}
where $\eta_c$ is the obliquity of Cassini state 1, $I$ is the inclination of Mercury's orbital plane with respect to the Laplace plane and $\mu_{LP}$ is the precession rate of the orbital plane about the Laplace plane normal \citep[see also ][]{2006Icar..181..327Y}. As the orbital characteristics of Mercury are accurately known eq.(\ref{EQ:PealeExpCore}) shows that the observation of the obliquity, libration, and degree-two gravitational coefficients is sufficient to determine $C_m/C$, the ratio of the mantle to the polar  moment of inertia. This reasoning is only valid if Mercury is in or close to Cassini state 1.
The spin axis orientation determined by \citet{2007Sci...316..710M} lies in the plane defined by the orbit pole and the Laplace pole of \citet{2006Icar..181..327Y} as required for a Cassini state.
Recent observations by \citet{2007Sci...316..710M} confirm the validity of this hypothesis with a measured obliquity value of about 2.11 arcmin, which lies precisely within the theoretically predicted range \citep{1988merc.book..461P,2002M&PS...37.1269P}.

The main goal of the BepiColombo rotation experiment is to obtain insight into Mercury's deep interior from observations of variations of Mercury's uniform rotation and precise measurements of its orientation in space. In particular this experiment uses observations of the 88-day forced libration, the obliquity, and the degree-two coefficients of the gravity field of Mercury \citep{1976Icar...28..459P,2002M&PS...37.1269P}. The gravity coefficients will be determined from the radio science experiment MORE (Mercury Orbiter Radio-science Experiment) \citep{2009Acta...65....1I} while the obliquity and librations will be determined from the comparison of observed target surface positions at different times.

Using the on-board star tracker to precisely determine the MPO's orientation and using the position determined through the radio-science experiment MORE, the location in inertial space of each image taken from the planetary surface by the SIMBIO-SYS HRIC (Spectrometer and Imagers for MPO BepiColombo Integrated Observatory System High Resolution Imaging Channel, \citep{2010P&SS...58..125T})  can be determined \citep{2009Acta...65....1I,2001P&SS...49.1579M}, and changes in landmark positions can be measured. 

The realistic estimation of the attainable accuracy of both the obliquity and libration amplitude determination in the frame of the BepiColombo mission is the main focus of this work. Previous simulations by \citet{2001P&SS...49.1579M} and \citet{2004P&SS...52..727J} used the best knowledge of the BepiColombo mission available at that time. Today many mission parameters potentially influencing the rotation experiment have changed and new constraints arisen. Additionally we also take into account the main gravitational perturbations the MPO will be subjected to.

Such an analysis requires the use of an accurate model of the rotation of Mercury, which takes into account other longitudinal librations besides the main 88-day libration due to planetary perturbations on Mercury's orbit. As the long-period longitudinal libration amplitudes depend on the same internal structure quantities as the annual libration \citep{2008CeMDA.101..141D,2009Icar..199....1P} we study the possibility of an independent determination of each amplitude in order to better constrain Mercury's interior structure.   

The paper is organised in the following way: In the following section the different models used to perform this study will be described. First the Mercury rotation model used (sec. \ref{subsec:MercRotModel}), then the MPO orbit model (sec. \ref{subsec:MPOOrbMod}), the observable and the method used to simulate the rotation experiment (sec. \ref{subsec:Observable}) and the considered error models (sec. \ref{subsec:ErrorModels}). We next discuss the possibility to perform repeated surface observations from the MPO in sec. \ref{sec:SurfaceObs}. In section \ref{sec:Results} we discuss the achievable precision levels on the annual and planetary librations as well as the obliquity. Finally some concluding remarks and perspectives are laid out in section \ref{sec:ConcPersp}.

\section{Method}
\label{sec:Method}

In the frame of this work, we create a simulated dataset consisting of the central position of potential surface images as a function of time as seen by the BepiColombo MPO. We consider an MPO orbit as defined by \citet{CREMA} (see table \ref{tab:allMPOorbitsV}) taking into account the perturbative effects of the low-level gravitational field coefficients $C_{20}$, $C_{22}$ and $C_{30}$ (see section \ref{subsec:MPOOrbMod}). By projecting the spacecraft orbit on the surface of the planet, the distribution of potential images on the surface is then determined (see section \ref{subsec:Observable}). Each potential image is subjected to a positioning error due to the uncertainties on the satellite position and orientation. The image-pairing process is represented by an additional error depending on the photographic resolution of the least well-resolved image and strict limits on the allowed surface solar illumination angle.  (see section \ref{subsec:ErrorModels})
The simulated dataset is then used to estimate the accuracy of the reconstruction of the orientation and rotational motion of the planet, as a function of the quantity of measurements made, the number of different targets considered and their locations on the surface of the planet in sections 3 and 4.

\subsection{Mercury Rotation Model used}
\label{subsec:MercRotModel}

\citet{2002M&PS...37.1269P} have shown that the differential equation governing Mercury's libration $\gamma$ can be expressed as

\begin{align}
\frac{d^2\gamma}{d^2 M}  & = - \alpha
\sum_{q}{G_{20q}(e)\sin\left[2\gamma+(1-q)M\right]} \; ,
\label{EQ:LibrationKaulaEq3}
\end{align}
where $M$ is the planetary mean anomaly and the $G_{20q}(e)$ are eccentricity functions of the type $G_{lpq}(e)$ as defined by \citet[(chap. 3)]{1966tsga.book.....K} and $\alpha = \frac{3}{2}\frac{(B - A)}{ C_m}$. 

The largest libration has a period equal to the period of the orbital motion of the planet around the Sun (87.969 days = 1 Mercury year) and can be obtained from eq.(\ref{EQ:LibrationKaulaEq3}) by keeping only those terms proportional to $\cos M$ or $\sin M$. This condition imposes that $q=0$ or $q=2$ in eq.(\ref{EQ:LibrationKaulaEq3}) yielding:

\begin{equation}
\gamma=\alpha\left(G_{200}-G_{202}\right) \, .
\label{EQ:AnLibAmpEq}
\end{equation}
In general Mercury's forced libration can be expressed as a series of sine terms representing the harmonics of the annual frequency. In order to describe Mercury's libration accurately enough for our simulation purposes, we need to take into account all harmonics with an amplitude larger than the expected precision of the BepiColombo libration experiment. It suffices to take the semi-annual term into account:

\begin{equation}
\gamma\left(t\right)=\gamma\sin M(t) + \gamma_{44}\sin 2M(t) \, ,
\label{EQ:LibrationAnnSemiAnn}
\end{equation}
where $\gamma_{44}$ is the semi-annual amplitude. Incidentally the semi-annual libration amplitude does not add an unknown to our system, as the amplitudes of the higher harmonics of the annual libration are related through a fixed coefficient to the annual libration amplitude. These coefficients, which we call $K$ for the first harmonic, only depend on the eccentricity of Mercury's orbit through the eccentricity functions. In the case of the semi-annual (44-day) libration we have: 

\begin{equation}
K=\frac{\gamma_{44}}{\gamma}  \frac{G_{20-1}-G_{203}}{4(G_{200}-G_{202})} \, ,
\label{EQ:K_depends_on_e}
\end{equation}
where the eccentricity function $G_{l,p,q}(e)$ only depends on the eccentricity of the Mercury's orbit. For Mercury's current eccentricity value of $e=0.2056$ this leads to a value of $K=-0.105455$. The semi-annual amplitude of Mercury's libration is of the order of  $10\%$ of the annual amplitude, and we can write:

\begin{equation}
\gamma\left(t\right)=\gamma \left(\sin M(t) + K \sin 2M(t) \right) \, .
\label{EQ:LibrationApprox}
\end{equation}
Through the same reasoning we obtain that all amplitudes of shorter period harmonics of the annual forced libration  will be smaller than $1$ arcsecond. As we will show in section \ref{sec:Results}, the accuracy needed on the libration amplitude to fulfil the \textit{minimum} mission requirements corresponds to $\sim 3$ arcseconds. Therefore there is no need to model the forced libration to a higher precision in the frame of this work.

Long period terms (from which $M$ is absent) can be obtained by averaging eq.(\ref{EQ:LibrationKaulaEq3}) with respect to $M$ over a whole revolution. We only have terms from which $M$ is absent if $q\equiv 1$. Since torques are small the angle $\gamma$ is considered very small and $\dot{\gamma}$ can be considered constant over one orbital period. Therefore, time-dependent forcing terms proportional to $\sin 2\gamma$ can be neglected in the right hand side of eq.(\ref{EQ:LibrationKaulaEq3}), reducing the equation to the harmonic oscillator equation. The resulting solution corresponds to the period of Mercury's long-term libration (or free libration, $\Pi_{\mathrm{Free}}$): 

\begin{equation} 
\Pi_{\mathrm{free}} = (2\alpha G_{201})^{-\frac{1}{2}} \frac{2\pi}{n}\; 
\label{eq:freelibperiod}
\end{equation} \citep{2004P&SS...52..727J}. For a value of $\alpha$\, of $3.06 \times 10^{-4}$ (based on the libration estimate obtained by \citet{2007Sci...316..710M}) and the expression of
$G_{201}=\frac{7}{2}e-\frac{123}{16}e^3$ this long-term period corresponds to $\sim 12$ Earth-years. 
Due to different dissipation processes inside the planet this libration should have damped out and no observable amplitude should have remained, since \citet{2005Icar..178....4P} has shown that the damping time is $<10^5$ years for all types of known excitations. This reasoning is valid except if a hitherto unknown process excites this libration.

\textit{Planetary induced librations} are caused through the same interaction mechanism as forced librations. The periodic torque arising from the interaction of a certain Solar System body with Mercury will cause a perturbation of Mercury's orbital motion, altering its position relative to the Sun and thus the solar torque acting on its equatorial bulge. This in turn will modulate the planet's libration motion and can be modelled as an additional libration of the same period as the perturbing forcing. The most important libration is caused by Jupiter and has a period of 11.86 years, coinciding with the time it takes for Jupiter to orbit the Sun once. The main libration caused by Venus has a period of 5.66 years corresponding to a 2:5 resonance between Venus' and Mercury's solar longitudes (orbital periods) \citep{2007Icar..187..365P, 2008CeMDA.101..141D}. 

\citet{2009Icar..199....1P} have shown that for the accepted value of $\alpha=3.06~\times~10^{-4}$ Mercury's libration will be dominated by two periods, the 88-day forced libration period and a long period libration forced at Jupiter's orbital period. The excitation mechanism at the basis for the large Jupiter induced libration amplitude is to be found in the proximity with the planet's free libration period. The independent measurement of its amplitude will pose an additional and independent constraint on  $\frac{B-A}{C_m}$ because its amplitude is highly sensitive to the free libration period.

To perform the Peale experiment Mercury needs to occupy Cassini state 1, as eq.(\ref{EQ:PealeExpTerm3}) is based on this hypothesis. The recent Earth-based radar observations by \citet{2007Sci...316..710M} have yielded a fairly accurate measurement of Mercury's obliquity, and indicate that this is the case.

\begin{table}[b]
	\centering
		\begin{tabular}{lcllcc}
		\hline
Body	&Symbol &	Period	&	Forcing argument	& \multicolumn{2}{c} {Normalized Amplitudes} 	\\
	&		& &	&	\citet{2009Icar..199....1P} & \citet{2009Icar....Dufey} \\ \hline
		Sun	& \astrosun &	43.98466 d	&	$2\lambda_0$ &	0.10223	&	0.11150 \\ 
		Sun	&	\astrosun & 87.96935 d	&	$\lambda_0$ &	1.00000	&	1.00000	\\ 
		Venus	&\venus&	5.66316 y	&	$2\lambda_0 - 5\lambda_V 	$&	0.09697	&	0.10691	\\
		Jupiter&\jupiter&	11.86295 y	&	$\lambda_J $ &	1.09339	&	1.2193	\\ \hline
		\end{tabular}
	\caption{A summary of the main libration amplitudes and associated periods, normalized relative to the annual libration amplitude. No other forced libration is modelled in this study}
	\label{tab:PlanetaryPerturbation}
\end{table}

If we define $\gamma_0$ as the initial angle $\gamma$ of the prime meridian at time $t_0=J2000.0$, and make use of the fact that the uniform rotation rate of the planet is equal to $\frac{3}{2}M$, the complete model of Mercury's rotation can be expressed as:

\begin{align}
\Theta \left(t\right)	&	=		\gamma + M t + \gamma(t) + \gamma_{\jupiter}(t) + \gamma_{\venus}(t) \,  ,
\label{EQ:RotModel}
\end{align}
where  $\gamma_{\jupiter}(t)$ and  $\gamma_{\venus}(t)$ represent the main Jupiter and Venus induced libration respectively. $\Theta$ as defined above is expressed in Mercury's non-rotating equatorial frame (Hermean mean equator of J2000.0). 
The detailed form of the planetary librations can be expressed as follows: 

\begin{align}
 \gamma_{\jupiter}\left(t\right) &	= \gamma_{\jupiter}\sin (\lambda_{\jupiter}+\varphi_{\jupiter})  \nonumber \\ 
 \gamma_{\venus}(t) & = \gamma_{\venus}\sin (2 \lambda_{0}-5\lambda_{\venus}+\varphi_{\venus}) \, ,
\label{EQ:PlanLibModel}
\end{align}
where $\lambda_i$ and $\varphi_i$ represent the planetary longitude and libration phase of planet $i$ and $\lambda_0$ is Mercury's longitude.

\subsection{MPO orbit model}
\label{subsec:MPOOrbMod}

In order to precisely model the trajectory of the BepiColombo MPO spacecraft around the planet we have to take into account several deviations from the perfect  motion. Since the planet cannot be considered as a spherically symmetric body, for example because of the equatorial bulge, various orbital perturbations arise. Hence, we include effects on the spacecraft orbit of the low degree gravitational field harmonics $C_{20}$, $C_{22}$, $C_{30}$. As the value of $C_{30}$ has not yet been determined for Mercury we use the following estimate: $C_{30}=0.1 C_{20}$, as is also done by ESA \citep{CREMA}. We have verified that using a value of opposite sign for $C_{30}$ does not impact the results of this study in a significant way.

The value used for the equatorial flattening of the gravitational potential is the MESSENGER flyby estimate from \citet{2010Icar..209...88S}, $C_{22}=0.81\pm0.08~\times~10^{-5}$. The value of the polar flattening of the gravitational potential, $C_{20}=6.0\pm2.0~10^{-5}$ is taken from \citet{1987Icar...71..337A} and based on the Mariner 10 mission flyby's. The value of $C_{20}$ recovered from the MESSENGER flybys, is considerably lower in magnitude than the value obtained from Mariner 10 tracking. However, it is not as well constrained as the equatorial ellipticity because the flyby trajectories were nearly in the planet's equatorial plane. Additionally the value by \citet{2010Icar..209...88S} is inconsistent with the Cassini state assumption. 
In order for $C_{20}$ to be fully consistent with the Cassini state equation when using the \citet{2010Icar..209...88S} value for $C_{22}$, its value should be taken inside the following interval $-5.6~\times~10^{-5} and -4.0~\times~10^{-5}$, which corresponds to the \textit{lower} part of the confidence interval given by \citet{1987Icar...71..337A}. We choose the value for $C_{20}$ to be $-4.9~\times~10^{-5}$ and for $C_{22}$ to be $0.81~\times~10^{-5}$, in our simulations.

We calculate the motion of the spacecraft numerically integrating the Lagrange equations which govern the variation of the osculating orbital elements \citep{1966tsga.book.....K}. This process assumes a trajectory influenced by Mercury's low level gravitational field only ($C_{20}$,$C_{22}$,$C_{30}$) and does not take into account for potential orbital manoeuvres or spacecraft wheel offloading.
We have tested that $C_{20}$,$C_{22}$ pairs at the boundaries of the expected range do not affect the results.

\subsection{Observable}
\label{subsec:Observable}

In order to perform the simulation of this experiment we need to express the position of any given image of the planetary surface in an inertial frame. This will then enable to determine the rotation parameters of the planet from comparing the image positions using landmarks in them. In this study this approach is simplified by comparing surface point positions directly.

A point on the planetary surface can be expressed in a Mercury body-fixed frame in spherical coordinates $(r, \phi, \lambda)$, where $r$ is the planetary radius, $\phi$ is the latitude and $\lambda$ is the longitude of the surface point. The spherical coordinates are defined with respect to the rotation axis \textit{z}, the \textit{x}-axis pointing in the direction of the planetary prime meridian at J2000 and the \textit{y}-axis in the equatorial plane perpendicular to the \textit{x} and \textit{z}-axis.  We here assume Mercury's shape to be spherical, to simplify our surface position calculations. This simplification amounts to the hypothesis that the shape of the planet is well known and does not represent an error source for the experiment. 
We express the position of a fixed point on the planetary surface in an inertial frame by performing a series of rotations. First a rotation from the rotating body-fixed frame to a Mercury non-rotating frame is performed.  We then perform a transition from an equator-based reference frame to an orbital-plane-based reference frame through the application of 3 rotations with respect to the axes \textit{z-x-z}, in order to take the planetary obliquity $\eta$  into account.
The first rotation about the \textit{z}-axis aligns the equatorial frame's \textit{x}-axis with Mercury's vernal point at J2000 . The second rotation about the \textit{x}-axis, then aligns the equatorial plane with the orbital plane, with an angle equal to the obliquity. The third rotation is an inverse of the first in order to align the \textit{x}-axis with Mercury's prime meridian at J2000. This yields:

\begin{align}
\left(\begin{array}{c}
x\\ y \\ z \end{array} \right)_{inertial} & = R_3\left(\varphi_{\eta}\right)R_1\left(-\eta \right) R_3\left(-\varphi_{\eta}\right)R_3(-\Theta)\left(\begin{array}{c} x\\ y \\ z  \end{array} \right)_{fixed} \, ,
\label{eq:ThothModelf90}
\end{align}

where $\varphi_{\eta}$ is the planetary longitude of Mercury's vernal point at the observation epoch,  $R_3(-\Theta)$  the rotation matrix with respect to the \textit{z}-axis, and $\Theta$ the planetary rotation angle defined in eq.(\ref{EQ:RotModel}). This yields the expression of a point on Mercury's surface in what is called here the \textit{Hermean Mean Ecliptic of J2000} frame which depends on the planetary rotation (rate) and obliquity, which are our main parameters of interest. The Hermean Mean Ecliptic frame (HME) is the frame in which we express our observable and perform all our calculations. It is the hermean analog to the Earth's Mean Ecliptic of J2000 frame.

In our simulation images are identified through their centre point expressed in the Hermean Mean Ecliptic frame. This does not mean that it is assumed that each centre point corresponds to a landmark. Instead this means that a landmark present in the image can be precisely positioned with respect to the image centre. This choice has no influence on the results of the rotation experiment analysis, as in reality images will be taken with existing knowledge of available targets. Not every surface point can be used as a landmark. The image correlation process that will be used to match images of a specific surface region, needs features that are sufficiently well resolved, like steep crater walls or albedo spots with very clearly defined borders. Additionally the solar incidence angle on the surface plays an important part in guaranteeing sufficient contrast and limited shadowing in order to successfully match two images. The reasons behind this will be discussed in section \ref{subsubsec:ImageCorrErr}. The images to be paired in this way do not need to be centred on the same surface spot, but only to have a sufficient amount of observed surface in common, which in turn has to contain enough \textit{adequate features} for the image correlation process. As no current images of Mercury's surface match the high resolution expected from the MPO High Resolution Imaging Channel (HRIC), we assume that adequate target regions are distributed evenly about the planetary surface.
The important quantities for our simulation are the timing of image acquisition and the volume of image data used for the parameter estimation, as we will show below, and not the precise location on the planet of the simulated targets used.

The above procedure reduces the process of \emph{generating} image pairs to the
determination of position differences quite simple:

\begin{align}
\vec{\Delta x}_{t_1 \, t_2} &	= R_3\left(\varphi_{\eta}\right)R_1\left(-\eta \right) R_3\left(-\varphi_{\eta}\right) \left[R_3\left(-\Theta(t_1) \right) \vec{x}_{t_1}-R_3\left(-\Theta(t_2) \right) \vec{x}_{t_2}\right] \, .
 \label{eq:ThothModel_DELTA_XYZ}
\end{align}
where the indices $t_1$ and $t_2$ correspond to the epochs of the two images to be compared. This quantity represents the observable in our simulations. 

In order to realistically simulate the restitution of the rotation parameters from these simulated observables we apply observational errors on each individual measurement of the position of a surface point in inertial space.

\subsection{Error Models}
\label{subsec:ErrorModels}
\label{sec:ErrorModels}
The following error sources are considered:
\begin{itemize}
\item satellite attitude error (composed of the star-tracker pointing error and the lack of knowledge of the orientation of the camera with respect to the spacecraft bus),
\item satellite position error,
\item image timing error,
\item image correlation error.
\end{itemize}

The first three errors refer to the inaccuracy in the determination of the spacecraft's absolute state and orientation, the fourth is due to the image correlation process, mainly because of the difference in illumination, altitude and orientation between two photographs to be correlated.

\subsubsection{Attitude error}

The attitude error represents the misalignment of the camera with respect to the expected nadir pointing position due to an error in the attitude determination or control. It is composed of a star tracker component and a camera component. The error in the knowledge of the pointing of the star tracker with respect to a celestial reference frame is considered to be better than 2.5 arcsec \citep{2001P&SS...49.1579M}. The misalignment of the camera with respect to the spacecraft optical bench (i.e. due to thermal stresses ) is considered to be below 2.5 arcsec \citep{2001P&SS...49.1579M}.

Those errors create a shift of the surface image effectively captured by the camera with respect to the centre of the nominal nadir image. We model the error through a randomly generated shift vector that gives the direction and the modulus of the misalignment of the pointing direction with respect to the expected/nominal nadir position. For the purposes of our simulations we use a
uniform random distribution $U[0,2\pi]$ for the direction of the attitude error, and a normal random distribution $N[0,\sigma_x]$ to model its amplitude. The attitude error amplitude $\sigma_x$ has been set to a value of $5$ arcsec in this study, corresponding to a worst case scenario in which the camera pointing error is at its maximum.
This creates a cone of uncertainty centred on the nominal nadir (ground track) surface in which we randomly position the observed image centre.
To quantify the impact of different camera pointing error levels on the libration and obliquity estimation process we perform simulations with errors ranging from 2.5 to 5 arcseconds. The results of this comparison are presented in section \ref{sec:PointErrCompar}.

\subsubsection{Position error}
The satellite position error is due to a shift of the satellite centre of mass with respect to the expected position  and results in a surface image shift that contributes to the error on the estimation of the planetary orientation and rotation rate. As for the attitude error, we express it through a random error vector with a Gaussian distribution around the actual position of the spacecraft centre of mass. The variance of this error can be determined from the precision of the satellite position estimation process, which varies according to the availability of spacecraft tracking. The spacecraft position will be determined through the MORE radio-science experiment. We consider two different error magnitudes depending on the tracking possibilities. The first case corresponds to a very small spacecraft position error of the order of $20$cm (rms) and is applied whenever the BepiColombo MPO spacecraft is in view of and tracked by a DSN station on Earth. The second case corresponds to an error larger by 2 orders of magnitude ($\approx 10$m) and represents a worst case scenario whenever the spacecraft is not tracked from Earth \citep{2001P&SS...49.1579M,2009Acta...65....1I}. %  \textbf{along track error >> 20cm even when tracking, see Berlin discussions}

We can conveniently divide the error in terms of along-track, cross-track and radial components. The radial component of the spacecraft position will be a lot better constrained than the components inside the plane perpendicular to it (by about one order of magnitude typically for other space missions). Additionally a small error on the spacecraft altitude will be of limited impact to our purposes. It will not significantly alter the surface area observed by the camera, resulting in a very small pixel size variation only. A radial error of 10 meters made from an altitude of around $1500$km\ \footnote{The altitude of $1500$km coincides with the nominal apoherm altitude of the BepiColombo MPO of $1508$km. (see table \ref{tab:allMPOorbitsV})} results in a MPO HRIC pixel size variation of $\sim 0.1 \textrm{mm}$ and is thus negligible for our purposes\footnote{The BepiColombo HRIC has a physical sensors with $2048\times 2048$ pixels, and a total field of view (FOV) of $1.47\degr$.}.  By contrast an along-track or cross-track error will shift the camera field of view by the same amount on the planetary surface independently of the spacecraft altitude, for example a $10$\textrm{m}  error in the along-track/cross-track plane will result in the same shift on the surface position of the target observed. We can further simplify our reasoning because the MPO orbit is quasi polar and the orbital period of the MPO is much smaller than Mercury's rotation, meaning that the spacecraft track is almost parallel to meridians. The along-track error corresponds almost exclusively to a shift of the image in latitude and the cross-track error to a longitudinal shift. The two error components are expected to be distributed normally, to be of the same order of magnitude and be independent of each other. 
The Earth-based tracking of the BepiColombo mission is expected to be performed by only one tracking station. We have considered the MPO visibility from the ESA Cebreros station. 

Therefore we apply a normally distributed longitudinal and latitudinal shift to each image centre with a magnitude of either $20$\textrm{cm} or $10$\textrm{m} depending on the visibility conditions of the spacecraft from the ESA Cebreros tracking station. %\textbf{cite why cebreros expected ... see berlin meeting}
In order to assess the impact of a lack of Earth-based tracking at the times where the surface is observed, a worst case with a constant spacecraft positioning error of $10$m has been considered, and is discussed in sec.\ref{sec:PointErrCompar}.

\subsubsection{Time tagging error}
Each image is tagged with its capture epoch. The time tagging error indicates a discrete error due to the wrong time clock reference on board the spacecraft. It is possible that the on-board clock rate changes with time by an unknown amount in the time between two images to be compared. This error step is random and with its value is associated a spacecraft position error vector, defined as the product between the velocity vector of the spacecraft per the step time. This means that the shifting is directed along track.

As the ellipticity of the MPO orbit is low we can approximate the MPO periherm velocity by its mean velocity:

\begin{equation}
\bar{v}=\sqrt{\frac{GM_{\mercury}}{a}} \, ,
\end{equation}
where $a$ is the semimajor axis of the MPO. Using $GM_{\mercury}=22032.09$ km$^3$s$^{-2}$ this yields a mean velocity of $2.54$km/s for the nominal semimajor axis value of $a=3393.7$km. Over 1 millisecond this corresponds to a displacement of $2.54$ meters which can be almost fully attributed to the along track component. Therefore all errors can be neglected if the on-board clock can be synchronised to better than $1/10$ of a millisecond (corresponding to below $25$cm). This level of clock precision is easily achieved, and therefore this error source has been ignored in our simulations.

\subsubsection{Image correlation error}
\label{subsubsec:ImageCorrErr}
\label{par:imagecorr_illumcond}
\label{par:ExplainNominalObservationCosntraints}
The image correlation error is related to the process of correlation needed to match images. This error depends on the difference between the respective image illumination and resolution (altitude), as well as on the presence of suitable surface features in the observed region. As said previously, no high resolution photographic data presently exists of Mercury's surface. For the purpose of this simulation we therefore postulate that suitable surface features are evenly distributed on the planetary surface, and sufficiently abundant for the experiment to be possible. 

An important part of Mercury's surface is covered with \emph{albedo spots}, small features with a very different albedo than their surroundings. As the albedo contrast tends to disappear for a very high solar incidence angle (Sun close to the zenith), the loss of albedo features and overall contrast render the image matching techniques all but useless. The same can be said of the usual topographic image matching. When correlating images of the planetary surface of Mercury we have two main characteristics of the surface to turn to: sharp topographic features (like craters and faults), and albedo spots. The rims of craters or potential large faults are very useful to precisely align overlapping images with each other, given that they are finely enough resolved. 
The main limiting factor for this is the crater morphology itself, over which we do not have any control, and the crater illumination. If the Sun angle is too grazing (low incidence angle) or too zenithal (high incidence angle) we will either have very long shadows or completely absent shadows. Without shadow it is impossible to correctly position the features on the surface.

\citet{Jorda..and..Thomas..00} simulated the pattern matching of albedo features and craters for a Mercury orbiter and concluded that sub-pixel accuracy can be achieved, if images obtained at phase angles $<5 \degr$ and $>55\degr$ are excluded (for albedo spots) and if the difference between the phase angles of the images compared/matched are smaller than 35$\degr$ (for craters). In our work we use their surface illumination criteria to select favourable targets and associated images, while guaranteeing pixel-size image matching errors at most. Therefore in this study we consider a standard 1-pixel error for each image pairing, scaled to the pixel size of the lowest resolution image. Based on the pixel size of the HRIC camera this represents an error of $2.5$ as, or $5$ to $18$ meters on the surface for MPO altitudes between $400$ and $1500$km.

All these illumination constraints are to our knowledge not fixed precisely in a definitive way, but are to be understood as worst cases. A more detailed study by \citet{BepiPatternMatch} specifically tailored to the BepiColombo mission using lunar images as well as synthetic images reached similar conclusions as \citet{Jorda..and..Thomas..00}. 

A recent experiment using image correlation to measure the libration of Epimetheus by \citet{2009Icar..204..254T} achieves sub-pixel accuracy around $0.5$ pixels. No limitations on the surface illuminations are made, as it is the movement of the shape of the body that is \emph{tracked} and not individual details at its surface. If such a sub-pixel level can be reached for the BepiColmombo rotation experiment, the level of uncertainty on the libration and obliquity estimation could be further reduced. The image correlation error has the same effect as adding an additional camera misalignment. Therefore a detailed study of the effect of different camera pointing errors (performed in sec.(\ref{sec:PointErrCompar})) can be used to quantify the effect of a reduced image correlation error.

It should be noted that the worst case image correlation error level of $1$ pixel is as large as the pointing errors best case.

\subsubsection*{Illumination constraints}
The following nominal illumination constraints are used in this study:
\begin{itemize}
\item Minimum solar elevation angle on the observed surface region: $ 35\degr$\, , 
\item Maximum solar elevation angle on the observed surface region: $ 85\degr$\, ,
\item Maximum difference in illumination angle between two images, to be able to compare them successfully: $35 \degr$ \, .
\end{itemize}
All images used have to satisfy these conditions in order to be paired and used
in the libration and obliquity estimation process.

\subsection{Required precision level}

What level of error on the obliquity and libration amplitudes is acceptable if we are to reach the BepiColombo rotation experiment mission objectives? The targeted precision for the $\frac{C_m}{C}$ and  $\frac{C}{MR^2}$ ratios stated in the BepiColombo mission goals  \citep[see][]{2001P&SS...49.1579M} is $\Delta(\frac{C_m}{C})< 10\%$ and  $\Delta(\frac{C}{MR^2})< 1\%$. 
By using eq.(\ref{EQ:PealeExpTerm2}) and eq.(\ref{EQ:PealeExpTerm3}) defined earlier we can not only relate the errors on the planetary parameters $C_{20}$, $C_{22}$, $\eta $ and $\gamma $ to the resulting estimation of $\frac{C}{MR^2}$ and $\frac{C_m}{C}$  but also express upper precision limits on the estimation of the rotation and orientation parameters posed by the mission requirements. Transforming those equations to isolate $\eta$ and $\gamma$ we get:

\begin{align}
\eta & =  \frac{ \sin I }{\cos I + \frac{n}{\mu_{LP}} \frac{M_{\mercury}R^2}{C} \left[\frac{ - C_{20}}{(1-e^2)^{\frac{3}{2}}} + 2C_{22}\left(\frac{7}{2}e-\frac{123}{16}e^3\right)\right]}   \, , \\ 
\gamma & =  \frac{3}{2}\frac{C}{C_m}\left(1 - 11e^2 +\frac{959}{48} e^4 + \dots \right)\frac{B-A}{M_{\mercury}R^2}\frac{M_{\mercury}R^2}{C} \, .
\label{eq:PealExp2i}
\end{align}
Using these relations we derive expressions for the upper error limits on both obliquity and annual libration amplitude that guarantee a successful experiment. We consider Mercury's orbital eccentricity $e$ and Mercury's orbital plane inclination with respect to the Laplace plane $I$ as error-free known quantities. Since the errors on the gravitational spherical harmonic coefficients $C_{20}$ and $C_{22}$ are expected to be determined with relative accuracy better than $10^{-3}$ \citep{2009Acta...65....1I}, the dominating factor limiting the precise determination of both quantities is the knowledge on the obliquity and libration amplitude value.  With respect to the errors on the obliquity and libration amplitudes, the latter can be expressed in terms of errors on $C/MR^2$ and $C_m/C$ as:

\begin{align}
\Delta \eta & =  \eta\Delta\left(\frac{C}{M_{\mercury}R^2}\right)\frac{M_{\mercury}R^2}{C} \, ,\\
\Delta \gamma & =  \gamma \left[ \Delta(\frac{C_m}{C} ) \frac{C}{C_m} - \frac{\Delta \eta}{\eta}\right]\, .
\label{eq:maxerr_OBL_LIB}
\end{align}

\citet{2009Icar..201...12R} have calculated interior structure models of Mercury for several plausible chemical compositions of the core and the mantle. For all models they have computed the associated $C/MR^2$ and $C_m/C$ values and thereby constrained their plausible value to the following intervals:
\begin{equation}
0.325 \leq \frac{C}{MR^2} \leq 0.37 \, , \quad 0.3 \leq \frac{C_m}{C} \leq 0.6 \, .
\end{equation}
These values together combined with the current best estimates for $\gamma$ and $\eta$ by \citet{2007Sci...316..710M} ($\eta=2.11~\pm~0.1~\mathrm{arcmin} \, , \gamma=35.8~\pm~2~\mathrm{arcsec}$), yield the following worst case estimate:

\begin{align}
\Delta \eta & \leq   1.0 \rm{as} \, ,\nonumber \\
\Delta \gamma & \leq  3.1 \rm{as} \, .
\label{eq:maxerr_OBL_LIB_Num}
\end{align} 
Increasing the precision requirement on $\frac{C_m}{C}$ to $5\%$ instead of $10\%$ increases the expected precision level on the libration amplitude to $\gamma \leq 1.4$ \textit{arcsec} (keeping the maximum error constraint on the obliquity identical).

\section{Surface observability}
\label{sec:SurfaceObs}

In this section we describe the influence of mission design constraints on the surface observability (useful surface for the rotation experiment) expressed in terms of surface coverage. Which parts of the surface can be observed at a given mean anomaly of Mercury depends on the orbital characteristics of the MPO. We here use the orbital elements defined by the European Space Agency as the BepiColombo baseline scenario (see table \ref{tab:MPOorbit-ESAS-N}). This orbit with a polar configuration coupled to the low altitude represents a favourable case for the photographic coverage of the planetary surface and so for the rotation experiment. 

\begin{table}[htb]
	\centering
	\begin{tabular}{c|c|c}
		Parameter				&	Symbol	& Value 					\\ \hline 
		semimajor axis 			& $a$ 		& $3393.7 \ km$ 	\\
		periherm altitude 		& $a(1-e)-R_{\mercury}$	& $400 \ km$			\\
		apoherm altitude 		& $a(1+e)-R_{\mercury}$	& $1508 \ km$			\\
		eccentricity 			& $e$ 				& $0.163244$ 			\\
		inclination 			& $i$ 				& $90.0 \degr$			\\
		argument of periherm 	& $\omega $ & $16 \degr$ 		\\
		longitude of ascending node & $\Omega $ & $67.7\degr$ 	\\
	\end{tabular}
  \caption{Initial BepiColombo MPO nominal orbital elements in the Mercury equatorial system on 21st July 2020, 0:00 UT as defined by \citet{CREMA}. $R_{\mercury}=2439.7$km }
	\label{tab:MPOorbit-ESAS-N}
	\label{tab:allMPOorbitsV}
\end{table}

This mission scenario proposes a mission launch date around July 21st 2014 resulting in an orbit insertion around Mercury on May 21st 2020 and reaching the final MPO orbit described above on July 21st 2020. Science operations are expected to begin after a one month commissioning phase around Aug 21st 2020. In this study we start the simulation of the orbit on the 21st of July but start the surface observations only on the 21st of August, to account for the effect of the small orbital evolution over the month of the commissioning phase. 

Previous studies have shown the importance of the spatial orientation of the MPO orbital plane for the libration and obliquity determination. This is due to the 3:2 spin-orbit resonance of Mercury and the fact that the MPO polar orbit's plane will not precess. An orbital plane with $\beta=0\degr$, where $\beta$ represents the angle between the Sun-Mercury periapsis line and the orbital plane at Mercury perihelion,  has been shown by \citet{2004P&SS...52..727J} and \citet{2006AcAau..58..236S} to be the most favourable configuration for the
obliquity and libration determination.  Our simulations confirm this and therefore we only consider $\beta=0\degr$ in this study.

Due to the proximity of Mercury to the Sun the BepiColombo MPO will be subjected to intense solar radiation. This poses a certain number of thermal issues, one of which in particular heavily impacts the spacecraft operations (Benkhoff, \textit{pers. comm.)}. During certain periods of Mercury's orbit around the Sun the solar radiation on the solar panels will cause them to overheat. At the basis of this lies the maximum allowable temperature of the solar panels, which cannot be crossed under penalty of permanent solar panel degradation. To limit the temperature of the panels their inclination with respect to the incoming solar radiation has to be modulated as a function of the Mercury true anomaly. Ironically it is when Mercury and the spacecraft will be closest to the Sun, when the most energy is available for potential harvest, that the situation will be critical. 
The enhanced solar radiation available at perihelion could be insufficient to offset the loss of power due to the larger solar panel inclination than at further distances of Mercury to the Sun. The power available to the MPO will then be set equal to the bare minimum needed to keep the MPO \emph{alive} (thermal management, attitude control, minimal communications), and science operations will not be possible. We have implemented these science blackout periods for the MPO as additional constraints in our simulations.

The primary \emph{science blackout windows} considered are situated between $20\degr$ to $50\degr$ and $310\degr$ to $340\degr$ of Mercury true anomaly. These windows are symmetrically disposed around the Mercury perihelion, due to the fact that the nominal MPO orbit has its orbital plane aligned with the planet-Sun direction at Mercury perihelion ($\beta=0\degr$). We study the presence of this exclusion zone as a worst case scenario, so as to predict its impact on the quality of the rotation and orientation estimation (librations $\&$ obliquity) in the frame of the BepiColombo rotation experiment.

In the current ESA outline for the MPO science operations it is not planned to make use of the SIMBIO-SYS HRIC in the first 180 mission days. The HRIC is crucial to the rotation experiment, as it is the only on-board camera with a pixel size smaller than the libration and obliquity signals we strive to estimate. The HRIC camera consists of a 2048x2048 pixels detector with a field of view (FOV) of $1.47\degr$, implying a pixel size of 5 meters at an altitude of 400 km and 19 meters from 1500 km. In order to evaluate the impact of a 180 day delay in HRIC operations on the feasibility of the Rotation Experiment we will also study short mission durations of only 180 days (the nominal mission being 360 days). This will be discussed in section \ref{sec:Results}.

\subsection{Nominal orbit}
\label{par:NominalOrbit}
\label{subsec:NominalOrbit}

\begin{figure}[ht]
\centering	
\subfloat[Distribution of surface regions which are observable at least twice and from which at least one image comparison can be constructed. The colour scale represents the number of such image comparisons which are possible. The observations have been filtered according to an illumination constraint excluding all low solar elevations (below 35$\degr$), very high solar elevations (above 85$\degr$) and imposing that inside an image-pair the illumination angle difference cannot be greater than 35$\degr$.]{\label{fig:ImagePairMap} \includegraphics[width=125mm]{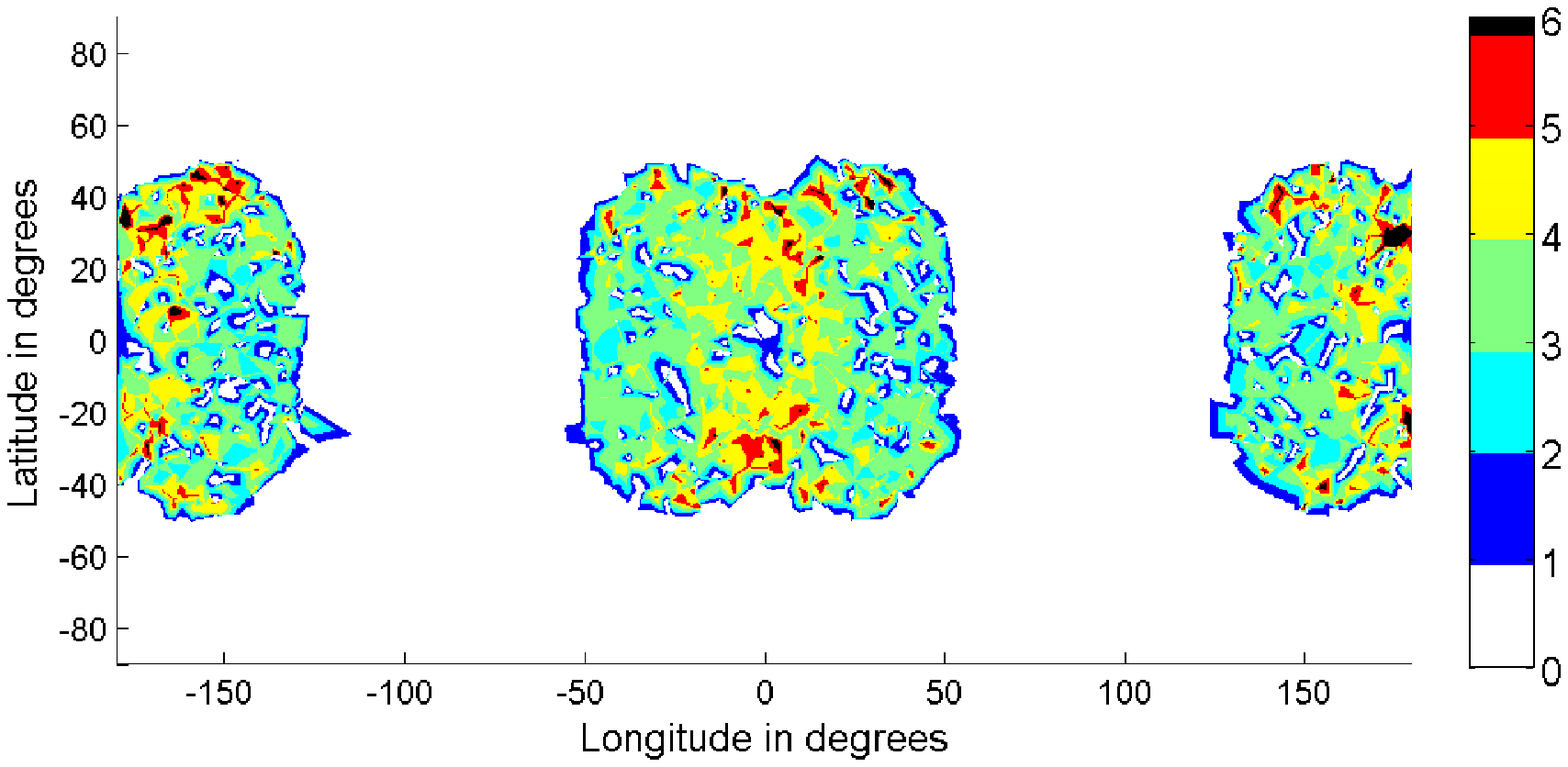}}

\subfloat[In addition to the same illumination constraints as fig.(\ref{fig:ImagePairMap}), the observations have been filtered according to a science exclusion zone between $20\degr$ to $50\degr$ and $310\degr$ to $340\degr$ of Mercury true anomaly. ]{
\label{fig:ImagePairMapMTA} \includegraphics[width=125mm]{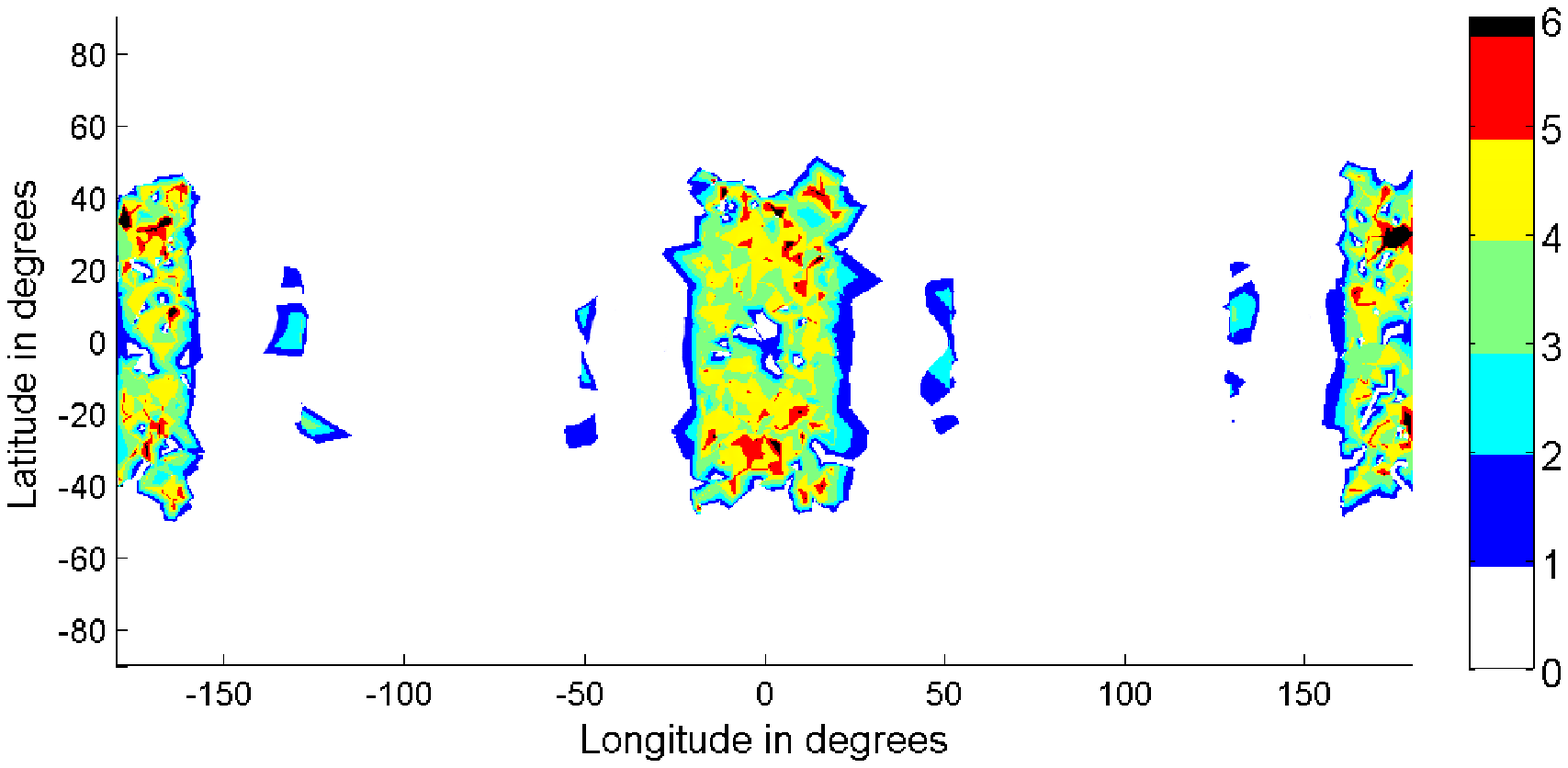}}
\caption{}
\label{fig:ImagePairMaps}
\end{figure}

In order to be able to determine libration and obliquity from camera images, landmarks of the planetary surface need to be flown over at least twice under favourable observation conditions. Therefore we determine the consequences of the various observation constraints imposed by the rotation experiment on the maximum number of image-pairs obtainable over the whole planetary surface after the nominal mission duration of 360 days and for the nominal orbit. Figure (\ref{fig:ImagePairMap}) shows how many image-pairs can be generated after 360 mission days (nominal) for the nominal ESA orbit under the previously defined nominal illumination constraints for each image-sized surface region on Mercury's surface. 

Observing at day-side is not a sufficient condition for having useful images. Because of the restrictions on the solar incidence (see section \ref{subsubsec:ImageCorrErr}) observations at latitudes North and South higher than 55$\degr$ where the Sun \emph{never} rises above 35$\degr$ in the sky are excluded for the determination of Mercury's rotation. We note that due to Mercury's very small obliquity the solar illumination angles are almost
symmetrical with respect to the equator. The lack of image pairs at high latitudes is clearly visible in fig.(\ref{fig:ImagePairMap}). Small patches of surface where the Sun is close to a zenithal position on overflight will also be excluded due to the lack of contrast arising from this situation. This corresponds to the illumination condition imposing that the Sun be lower than $85\deg$ in the local sky. 
The longitudinal distribution of the number of image pairs on the planetary surface as seen in figure (\ref{fig:ImagePairMap}) can be partly explained by Mercury's 3:2 spin-orbit resonance and the choice of a (quasi) polar MPO orbit. 
Taking into account that the plane of the polar orbit of BepiColombo's MPO will hardly precess, there are 12 possible overflights of a given meridian in the nominal mission duration of one terrestrial year ($\sim$ 2 Mercury days). This corresponds to one target overflight every half Mercury rotation or 29 days. At the times corresponding to these intersections, we need to check the illumination conditions. Useful images can only be obtained when the illumination conditions on the surface are favourable. Since Mercury's obliquity is close to zero (2.11 \rm{arcmin}, \citet{2007Sci...316..710M}) the illumination conditions can be considered as controlled by the difference between the planetary rotation angle and the true anomaly. 
There will be (on average over a range of longitudes) 6 opportunities when the surface will be on the day-side (the 6 others happening at night). Because of the 3:2 spin-orbit resonance, the orbital configuration of Mercury repeats exactly every Mercury day (176 days). Therefore, for each observed target on the surface there are two groups of (on average) 3 images, which are identically illuminated and observed at the \textit{same} Mercury true/mean anomaly. Since the libration in longitude is a function of the planetary mean anomaly, correlating images taken exactly one Mercury day apart yields no information on the annual libration amplitude $\gamma$.

The image pair maps in fig.(\ref{fig:ImagePairMap}) do not contain image pairs made at an identical planetary anomaly, nor images with illumination conditions differing by more than $35\deg$ of solar incidence, hence most of the correctly illuminated planetary surface is observed less than $6$ times yielding between $1$ and $5$ image-pairs. Certain longitudes will be \emph{oversampled} with respect to others because mission durations are not integer multiples of one Mercury day.

Although the above reasoning is made for the case of a Keplerian orbit without any secular or periodic variations the image repeatability is changed only slightly by inserting the perturbative effects of the low-degree gravity field harmonics. Nonetheless, as we have mentioned earlier a Keplerian orbit is not a good approximation since the perturbing effects of the low-order gravity field harmonics will be \textit{quite large} relative to the signal of the rotation experiment parameters. Therefore, only on average after 176 days the MPO will fly over the same surface regions and the whole set of possible observations will take 176 days or longer to fill (in steps of half Mercury rotation periods, corresponding to $\sim 29$ days, because the orbit plane does not precess and the planetary rotation unfolds \textit{under} the satellite).

In contrast to previous studies we do not consider a \emph{camera offset} capability in order to increase the number of observations of a given target / surface region. Simulations by \citet{2006AcAau..58..236S} have shown that a small camera offset capability yields more precise libration and obliquity estimates for an identical number of observations, because more favourable target observations and otherwise impossible image-pairs can be obtained. At the present time the only camera offset possible for the BepiColombo MPO is through off-nadir pointing manoeuvres. Their amplitude is limited as the MPO's payload risks overheating if not closely aligned with the spacecraft nadir. It should be noted that the addition of such a capability adds additional uncertainty on the spacecraft pointing direction and therefore the advantage of \emph{more} image-pairs is not entirely expected to benefit the libration and obliquity estimation, unless it is completely impossible to obtain satisfying image pairs without an off-nadir pointing . As there are currently no high resolution maps of Mercury's surface available this cannot be excluded, but seems highly unlikely. The results we present here represent a worst case scenario in which no off-nadir pointing is performed. 

The fact that the image-pair opportunities exhibit a latitudinal symmetry (Northern and Southern hemisphere a both equally covered) (see fig.\ref{fig:ImagePairMap}) is due to the fact that the initial orbital configuration is optimised to assure equal coverage of the Northern and Southern hemispheres over the nominal mission lifetime taking into account the secular drift of the periherm. In the nominal ESA (ESAS) case the periherm will advance by about $33 \degr$ per year, and starting at $-16 \degr $South will reach $16 \degr$North after one year. By advance of the periherm we mean the fact that the periherm occurs earlier in the orbit path than it would have been expected.

\subsection{True Anomaly limitations}
\label{subsec:TrueAnLim}

Because of reasons of thermal management of the solar panels no science operations can be performed in two zones in true anomaly close to Mercury's perihelion (see above).
Fig.(\ref{fig:ImagePairMapMTA}) shows the effect of the addition of science exclusion zones between $20\degr$ to $50\degr$ and $310\degr$ to $340\degr$ of Mercury true anomaly, on the repeated surface observability. 
Comparing fig.(\ref{fig:ImagePairMapMTA}) with fig.(\ref{fig:ImagePairMap}) we notice two additional empty longitudinal bands situated between $20\degr$ and $50\degr$ of East and West longitude. The excluded true anomalies are directly related to the planetary longitudes affected due to the 3:2 spin-orbit resonance. Therefore we can say that if an exclusion zone contains the Mercury perihelion, the observations of the surface region centred on the prime meridian will be reduced or even completely vanish.
The main effect of such a science blackout zone is to further reduce the available surface in which to look for suitable image-pair candidates. In the following section we compare the effects of the presence of such a limitations on the libration and obliquity estimation process.

\section{Results for librations and obliquity}
\label{sec:Results}
In this section we investigate the quality of the estimation of the studied parameters (libration amplitudes and obliquity) for the nominal BepiColombo MPO orbit in order to establish the feasibility of the rotation experiment at the desired precision level. The following parameters are considered as limiting factors on the surface observability: the solar incidence on the surface seen by the camera and the solar incidence variation between two measurements of the same target, the mission duration, the number of observed targets and the presence and size of science blackout zones. We studied the effect of various combinations of these parameters on the  estimation quality of the obliquity and libration amplitude. We also studied the effect of limiting the observations made by the MPO above 1000 km because of the trade-off between the resolution and the amount of pictures \citep{2001P&SS...49.1579M,2004P&SS...52..727J}.  

The results discussed in this section are obtained using \emph{simulated} targets located inside the observable zones defined in the previous section. Using these filtered observations we determine the value of the observable (the position change of a given image pair), add noise to the observable according to the error models and calculate the partial derivatives of the observable defined in eq.(\ref{eq:ThothModel_DELTA_XYZ}) with respect to the four parameters studied here. This is followed by the estimation of the parameters through a least squares approach, solving the following system:

\begin{align}
p & = \textbf{A}^{T}\textbf{C}_{b}^{-1}b(\textbf{A}^{T}\textbf{C}_{b}^{-1}\textbf{A})^ {-1} \, , \\
\textbf{A}_{il}& =\frac{\partial\Delta\vec{x}_i}{\partial p_l} , \qquad \emph{for} \quad i=1..4 \quad \mathrm{and} \quad l=\mathrm{\eta, \gamma, \gamma_{\jupiter},\gamma_{\venus}} \, ,
\label{eq:axeqb2}
\end{align}
where \textbf{A} is a ($m \times n$) matrix of the partial derivatives of our observable with respect to the different parameters (with \textit{n} the number of parameters (n=4) and \textit{m} the number of observations), 
$\textbf{A}^{T}$ is the transpose of  \textbf{A}, $b$ is a ($\textit{m}\times\textit{1}$) vector containing the residuals of the observations, and $p$ ($\textit{n}\times\textit{1}$) is the common value taken by the parameters throughout all observations which is to be determined by our least squares approach. $\textbf{C}_b$ is a ($m \times m$) weight matrix that can be applied to the observations. Generally it is a diagonal matrix composed of the inverse a priori uncertainties on the observations (\textit{b}) is used. In our case $\textbf{C}_b$ is equal to the identity matrix as we do not give more weight to certain observations relative to others. $\textbf{A}^{T}\textbf{C}_{b}^{-1}\textbf{A}$ is a square ($n \times n$) matrix, and is called the normal matrix.

We present here the formal errors reached on each parameter for different sets of constraints as a function of either the number of surveyed targets or the mission duration. 

\subsection{Number of surface targets surveyed}

\begin{figure}[htb]
\includegraphics[width=140mm]{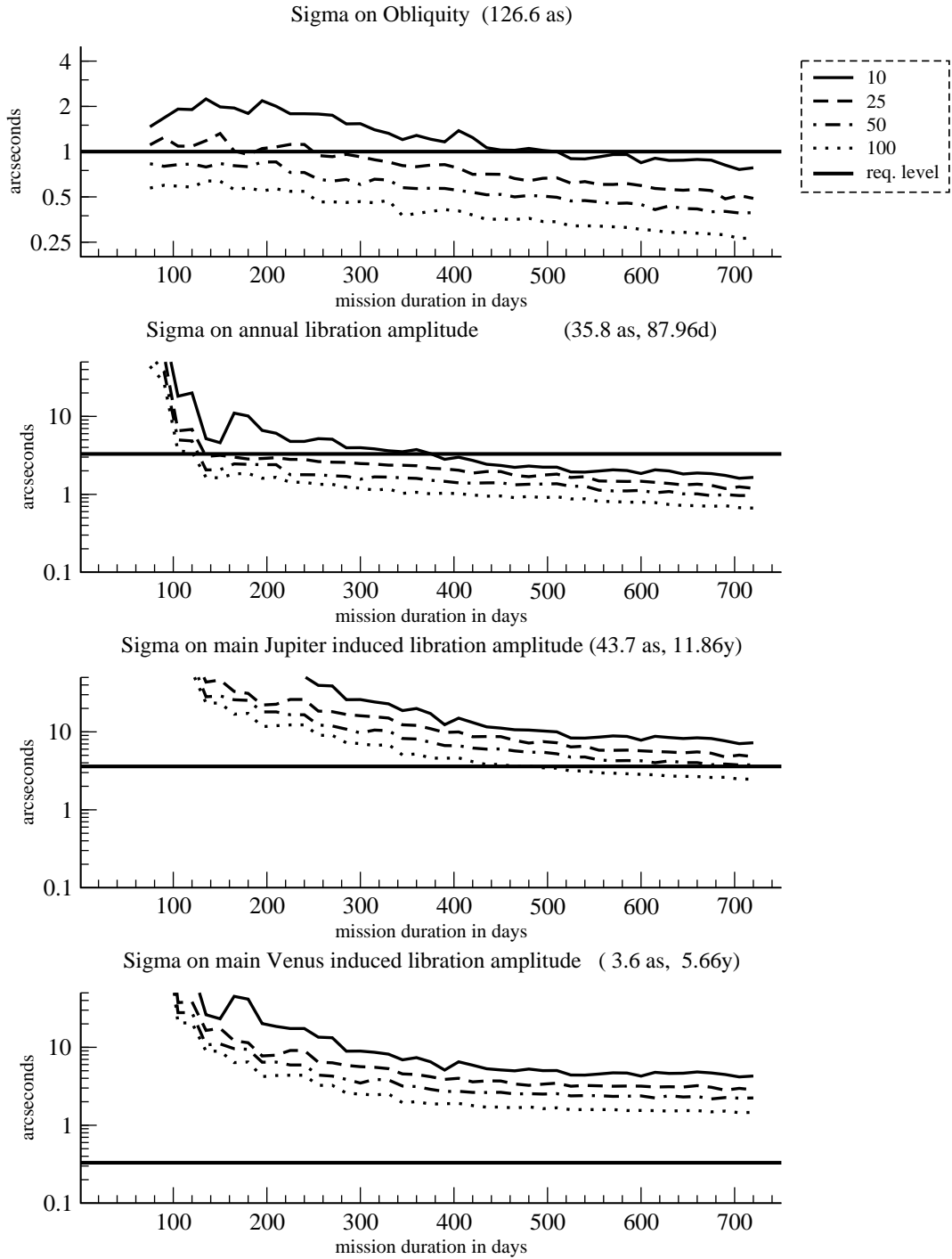}
\caption{From top to bottom the \textbf{formal error ($1\sigma$)} on the  four parameters we solve as a function of the mission duration. Each plot contains four lines representing each a different number of targets used to obtain the solution. The thick black line represents the error level required for a successful experiment. \textbf{The full, dashed, dash-dotted and dotted curves represent 10, 25, 50 and 100 targets used respectively.}}
\label{fig:TargNumCompar}
\end{figure}

Figure (\ref{fig:TargNumCompar}) presents the formal error on each of the four parameters as a function of the mission duration for different numbers of surveyed targets  (10, 25, 50, 100). The presence of a slow and steady decrease in formal error with mission duration for any constant number of surveyed surface targets is a strong indication that the data volume is not the bottleneck of the experiment. Indeed throughout the mission the surveyed targets will continue to yield additional image-pairs. The small decrease in formal errors when the target numbers increase is mainly due to the increase in data volume and the fact that the formal error is proportional to $\frac{1}{\sqrt{N}}$, where $N$ is the number of image-pairs. With a smaller number of targets the error on the rotation parameters increases but a sufficient precision could be obtained for a longer observation time.
% edited out ?
%can \textit{do the job} as well as a large number, if these are ideally located in terms of obliquity and libration signal and can be observed under optimal conditions. 
Nevertheless a large number of targets may be needed to reduce the error levels below the mission requirements. 

Results with the required precision (see eq.\ref{eq:maxerr_OBL_LIB_Num}) can be reached after 180 mission days already for the obliquity and annual libration amplitude, for 50 surveyed targets, as all different observation windows will already have been \textit{explored} after one Mercury day. For a small amplitude and  long period signal such as both long period planetary librations studied here, the observation time series has to be long enough to cover a significant part of the cycle so that the cycle can be determined accurately. 
The main error sources of the experiment will have to be reduced in order to reach the required precision on the planetary libration amplitudes, as more measurements and a longer mission duration do not suffice (see fig.\ref{fig:TargNumCompar}). 
A limited number of 25 targets suffices to reach an error level satisfying eq.(\ref{eq:maxerr_OBL_LIB_Num}) on both the annual libration and obliquity in the nominal 360 day mission.

All parameters except the planetary induced libration amplitudes can be solved for with a formal error smaller than the minimum mission requirement with 25 image-pairs after around 300 mission days as can be seen in fig.(\ref{fig:MTA_compar_fig}). The error on the Jupiter induced libration estimate is close to the mission requirements defined by eq.(\ref{eq:maxerr_OBL_LIB_Num}) after about 550 days. A doubling of the number of surveyed targets would enable to achieve a halving of the error value to barely reach the minimum requirement.

\subsection{Altitude limit}
\begin{figure}[htb]
\includegraphics[width=140mm]{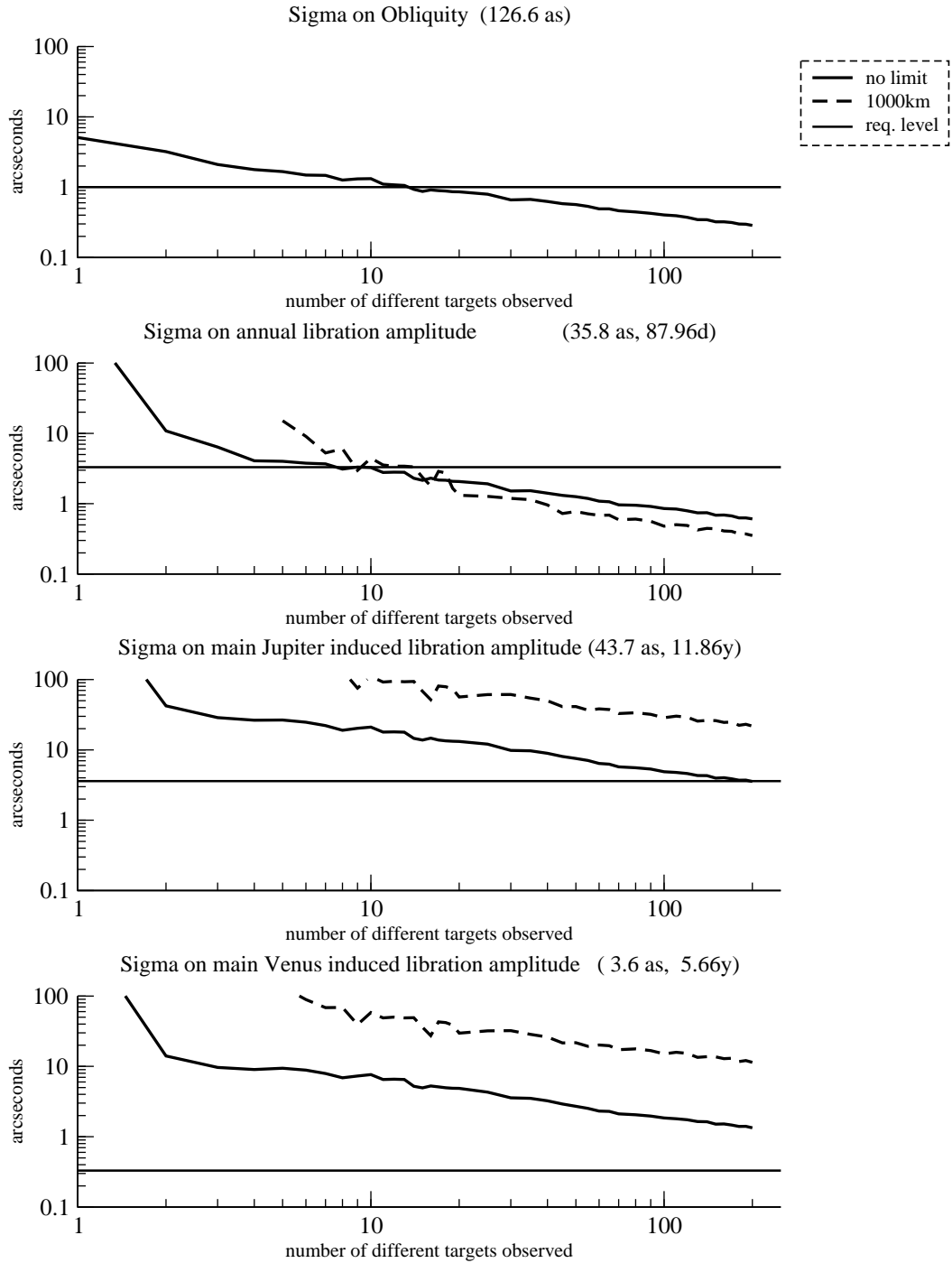}
\caption{From top to bottom the \textbf{formal error ($1\sigma$)} on the four parameters we solve for a 360 day mission for the nominal ESAS orbit. Each plot contains two curves representing a different upper altitude limit on the observations used in the results. The thick horizontal black line represents the error level required for a successful experiment. The black curve represents the \textit{altitude unlimited} scenario where images are not filtered based on the altitude from which they were taken. The dashed line corresponds to an altitude limitation of $1000$km on the observations used.}
\label{fig:AltComparFig}
\end{figure}

If measurements are restricted to 1000 km in altitude (reducing the data volume) all the parameters are less well solved for, except for the libration amplitude (fig.\ref{fig:AltComparFig}). The latter is best determined if observations are restricted to lower altitudes. This difference is due to a combination of the following two facts: The annual libration signal is strongest in the equatorial regions as it is a longitudinal signal, and the periherm of the MPO orbit is located close to the equator creating the possibility of very strong-signal
image pairs. If we allow images taken at higher altitudes we add image pairs of lower resolution spread over a wider latitude domain. Since we have not weighted the observations as a function of the latitude or altitude or any other characteristic (because this would have skewed the results of the other three parameters while favouring this one), the addition of lower resolution image pairs degrades the total solution, diluting the signal coming from high signal to noise image-pairs taken at low altitudes. The solution therefore gets \emph{noisier}. We don't see this for the other two libration amplitudes because of their very long period and therefore significantly smaller longitudinal displacement with respect to the annual libration movement. Their signal is \emph{weak enough} to benefit from the presence of higher altitude observations, which enable a greater number of image pairs. The fact that the obliquity does not benefit from the low altitude only approach is explained in the following way. Due to Mercury's 3:2 spin resonance and the fact that the MPO orbital plane stays fixed in space and is initially fixed to $\beta=0\deg$, two of the three favourable dayside surface observations (see section \ref{sec:SurfaceObs}) happen symmetrically around the Mercury aphelion, and correspond to \emph{low altitude} images taken from the MPO periherm orbit side. The third observation window is situated around Mercury perihelion and corresponds to \emph{high altitude} observations from the MPO apoherm side. Because both low altitude observations happen almost symmetrically about perihelion their comparison will yield no information on Mercury's obliquity as the planet will be almost identically oriented in space with respect to the MPO orbit. The third observation window provides a different viewing angle on the planet, allowing for image-pairs containing a strong obliquity signal. Therefore it is critical for the obliquity measurements not to limit the surface imaging to low spacecraft altitude.

\subsection{Science blackout zones}

\begin{figure}[htb]
\includegraphics[width=140mm]{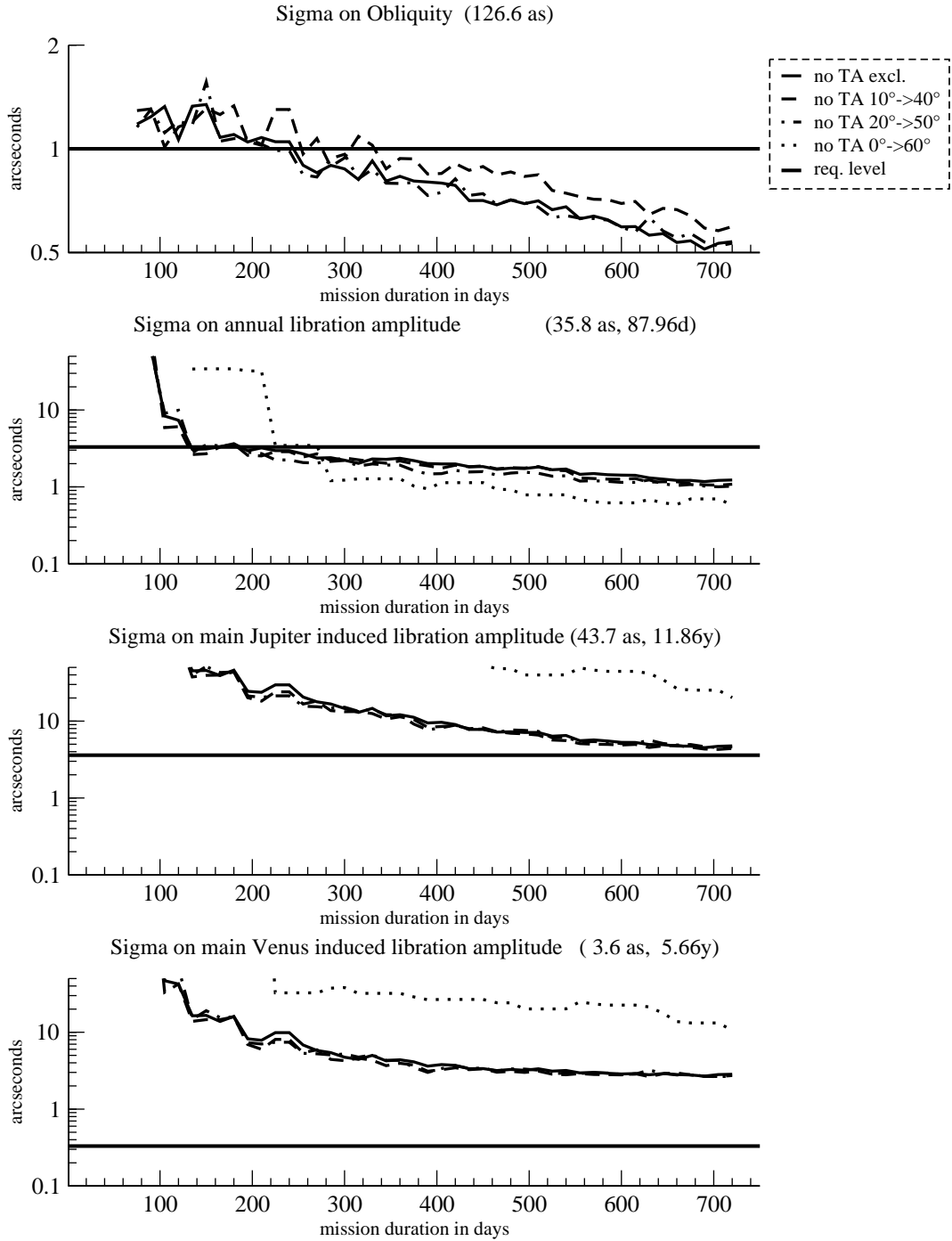}
\caption{From top to bottom the \textbf{formal error ($1\sigma$)} on the  four parameters we solve for a 360 day mission. Each plot contains four lines representing each a \textit{different Mercury true anomaly exclusion zone} constraining the observations used. The thick black line represents the error level required for a successful experiment. The black curve corresponds to the absence of a blackout zone. The dashed, dashdotted and dotted curves represent the different exclusion windows studied. For the obliquity, the dotted line is always above 2 arcsec.}
\label{fig:MTA_compar_fig}
\end{figure}

In fig.(\ref{fig:MTA_compar_fig}) we compare the results obtained for a fixed number of 25 surveyed targets as a function of the mission duration, for different \emph{blackout zones} and a case without any blackout zone. The zones considered correspond to the nominal case described earlier (see section \ref{sec:SurfaceObs}), one variant shifted towards Mercury perihelion, and one radical variant excluding all observations made between $300\deg$ and $60\deg$ of Mercury true anomaly. For this last zone we can see in fig.(\ref{fig:MTA_compar_fig}) some similarities with the altitude limitation as no observations can be made at and close to Mercury perihelion, such as no successful obliquity estimation.
The results for all \emph{blackout zones} allowing observations at perihelion are extremely similar and close to the result of \emph{unaffected} case. This shows us that the size of these exclusion zone is not the dominant factor in the degradation of the parameter estimation process, but the presence of observations made close to perihelion is. 
Surface observations between $-10\degr$ and $+10\degr$ of Mercury true anomaly seem to be very important for the achievement of the mission requirements as the \emph{$10\degr$ to $40\degr$} blackout zone is barely able to achieve the necessary precision in the nominal mission duration of 360 days.

\subsection{Illumination constraints}

\begin{figure}[htb]
\includegraphics[width=140mm]{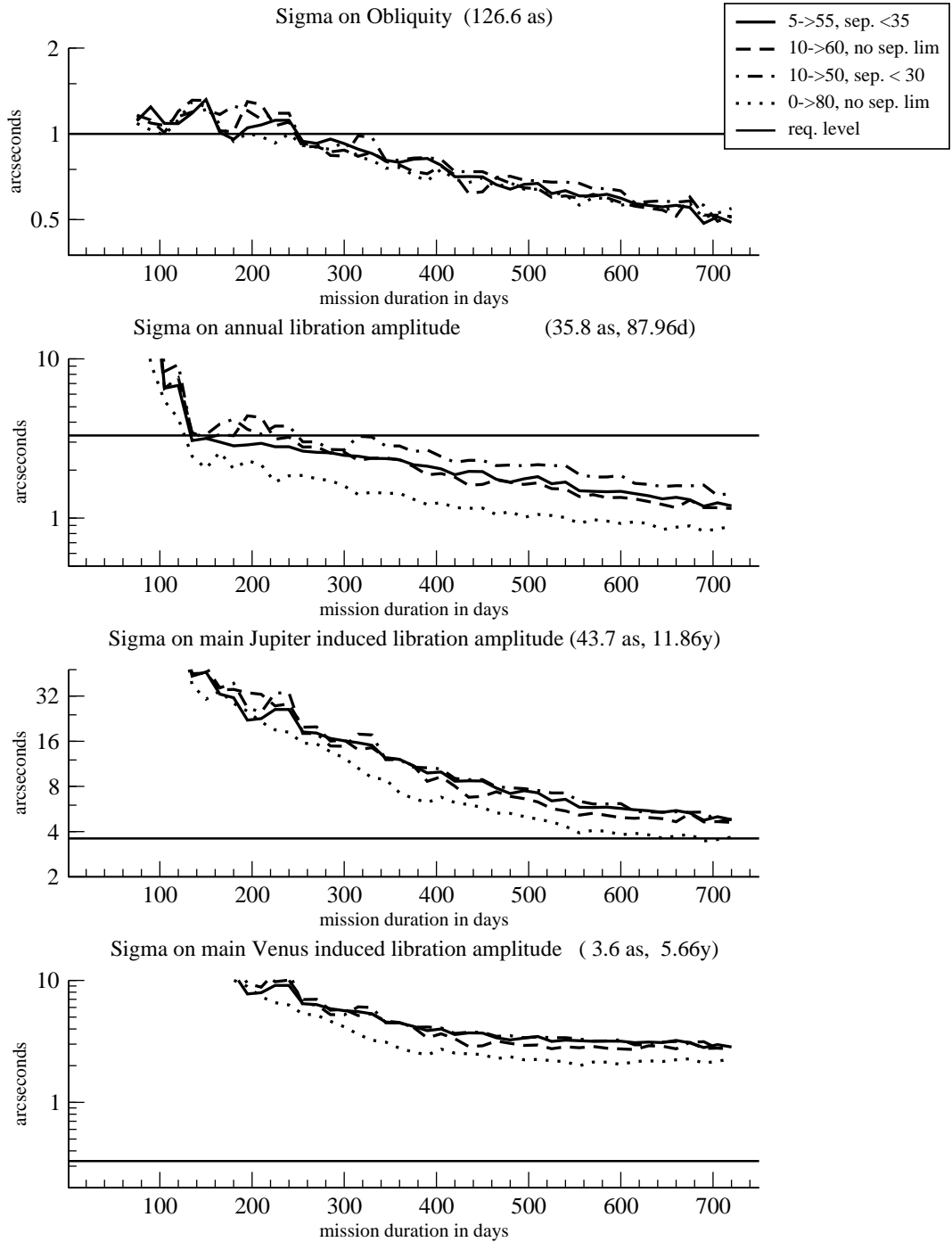}
\caption{From top to bottom the \textbf{formal error ($1\sigma$)} on the  four parameters we solve for a 360 day mission. Each plot contains four lines representing each \textit{different surface illumination conditions} constraining the observations used. The thick black line represents the error level required for a successful experiment. The black curve corresponds to the nominal illumination conditions. The dashed curve represents slighly relaxed conditions, without the illumination separation limitation for image pairs. The dashdotted curve corresponds to slightly more strict illumination conditions. Finally the dotted curve represents very lax illumination conditions without any separation limitation.}
\label{fig:IllCompFig}
\end{figure}

Figure \ref{fig:IllCompFig} shows the formal error as a function of the mission duration for a fixed number of 25 observed targets for the nominal illumination constraints versus relaxed and tightened illumination constraints. The first goal of this figure is to show the absence of influence of the difference in illumination angles between images forming an image pair (\emph{separation limit}) on the parameter estimation. The solid and dashed lines correspond to the nominal illumination conditions and the case of slightly relaxed conditions without the need for a maximum image illumination separation of $35\deg$ inside an image pair. Comparing them we cannot see significantly smaller errors for the dashed line. Comparing the nominal conditions to slightly stricter conditions represented by the dash-dotted line it can be seen that the stricter conditions enhance the errors significantly only for the libration amplitude. Finally the dotted line represents a quasi unlimited illumination scenario where only local solar elevations below $10\degr$ are being rejected. This allows for a hugely increased portion of the planetary surface, between the northern and southern $80\degr$ parallels and with an almost homogeneous longitudinal distribution, in which targets can be observed. Nevertheless for an identical number of surveyed targets, the nominal illumination conditions yield only about two to three times less favourable error bars on the parameter estimates. Again the annual libration amplitude is the most affected. We can clearly see that in all cases the formal errors decrease with mission duration. A satisfying precision level below 1 \rm{arcsecond} for the obliquity and 3 \rm{arcsecond} for the annual libration is reached after $\approx$\textbf{300} mission days for all sets of illumination constraints. 
From the perspective of the annual libration amplitude enough information is available to reach a precision smaller than defined by eq.(\ref{eq:maxerr_OBL_LIB_Num}) after approximately 250 mission days, for any of the illumination conditions studied. 

\subsection{Influence of the camera pointing error level}
\label{sec:PointErrCompar}

\begin{figure}[htb]
\includegraphics[width=140mm]{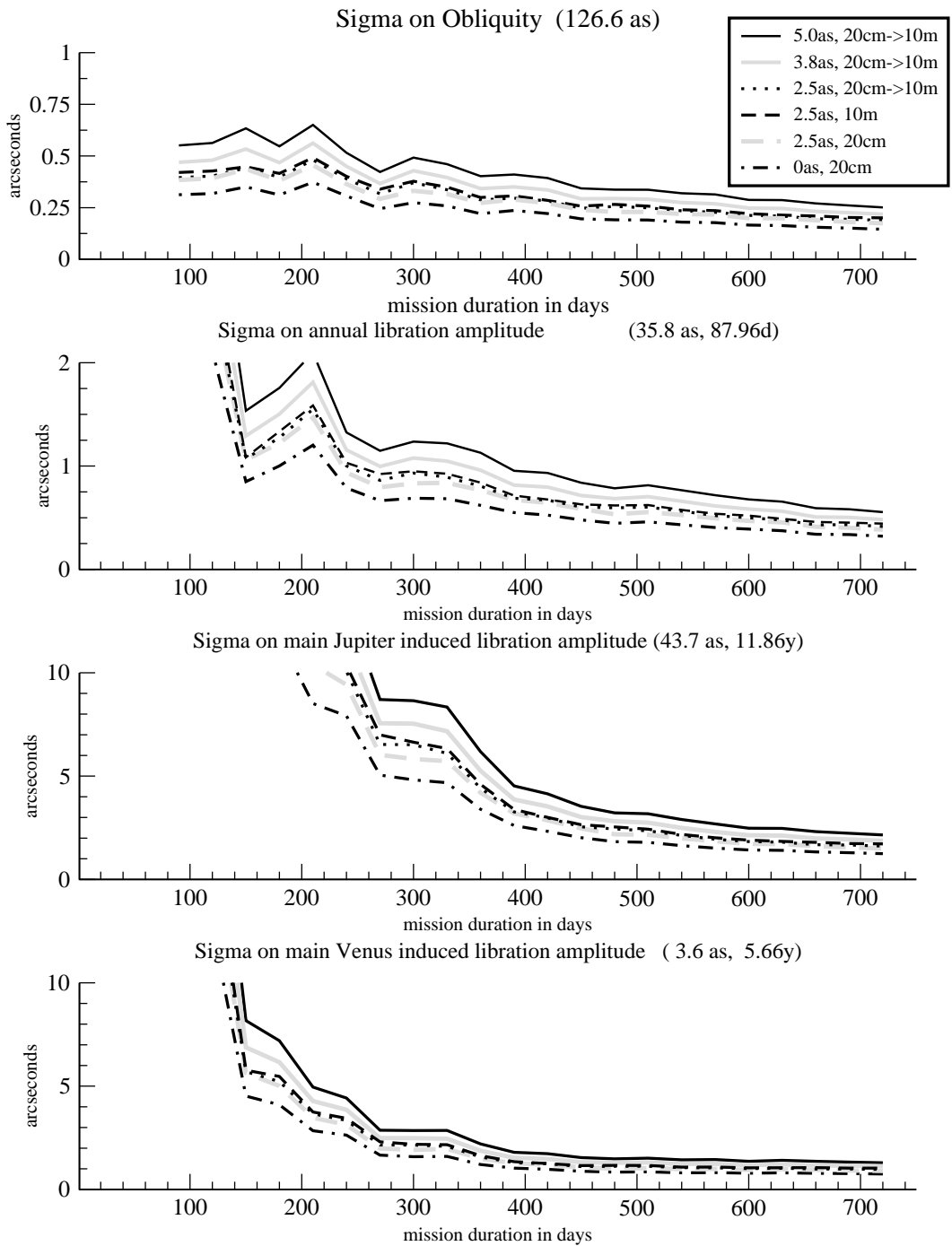}
\caption{From top to bottom the \textbf{formal error ($1\sigma$)} on the  four parameters we solve for a 360 day mission. Each plot contains four lines representing each \textbf{different camera pointing (in arcsec)  and spacecraft positioning (in m) error levels}. The black curve corresponds to the worst case pointing error scenario of $5$arcsec. The grey curve represents an intermediate scenario, with $3.8$arcsec of pointing error. The dotted curve corresponds to the most optimistic pointing error level of $2.5$arcsec.  The dashed curves stand for cases where the spacecraft positioning error is fixed to 10m (black) or 20cm (grey). Finally the dashdotted curve represents a case without pointing error, and with a constant 20cm position error.
}
\label{fig:PointErrComparFig}
\end{figure}

In order to gauge the impact of the camera pointing error level on the libration and obliquity estimation, we have simulated the experiment for error levels of $2.5$, $3.8$ and $5.0$ arcseconds. The comparison can be found in figure (\ref{fig:PointErrComparFig}) and shows that a reduction of the pointing error level by a factor of two reduces the error level on the obliquity and libration amplitudes by slightly less than a factor of two. 
If the pointing error can be reduced to its minimum level of $2.5$as the Jupiter induced libration \emph{could} be determined within the nominal mission duration ($360$ days) with an error level below the requirements to pose an additional constraint on the planetary interior ($<2.5$as). In general it can be said that a reduction of pointing error level from its worst case to its ideal case does not impact the order of magnitude with which the obliquity and libration amplitudes will be determined.

The impact of the MPO position error on the rotation experiment has been studied. The nominal scenario in which the MPO is tracked by one Earth-based groundstation has been compared to a situation with a constant $10$m position error. When tracked the MPO position error drops to $20$cm.  No impact on the final result from a constant $10$m error on the MPO position can be seen in fig.(\ref{fig:PointErrComparFig}).
%The radial component of the spacecraft  position will be a lot better constrained than the components inside the plane  perpendicular to it (by about one order of magnitude typically for other space  missions that include a laser altimeter). Additionally a small error on the spacecraft altitude will be of limited impact to our purposes. It will not  significantly alter the surface area observed by the camera. By contrast an along-track or cross-track error will shift the camera field of view by the same amount on the  planetary surface independently of the spacecraft altitude. We can further simplify our reasoning because  we are in the presence of a quasi polar orbit meaning that the spacecraft  track is almost equivalent to meridians. The along-track error corresponds  almost exclusively to a shift of the image in latitude and cross-track error to a  longitudinal shift. We can confidently apply the position error in terms of an  additional error on the image centre position in latitude and longitude. 

A position error of $10$m on the MPO position translates in a maximum error of $10$m on the planetary surface. By contrast a pointing error of $2.5$as from an altitude of $400$km corresponds to a surface displacement of $4.8$m. For $5.0$as this corresponds to  $9.6$m. From an altitude of $1500$km this represents an error of $18$m or $36$m respectively.

Two images composing an image pair are rarely both taken while the MPO can be tracked from Earth. Therefore the impact of a constant $10$m tracking error on the final result is limited. This is especially true for scenarios with larger MPO pointing error levels.

%\begin{figure}[htbp]
%\centering
%\includegraphics[width=90mm]{OrbitComparFig}
%\caption{From top to bottom the \textbf{formal error ($1\sigma$)} on the  four parameters we solve for a 360 day mission. Each plot contains four lines representing each \textbf{different values for Mercury's low level gravitational field coefficients}. The thick black line represents the error level required for a successful experiment. Each curve corresponds to a different set of values for  $C_{20}$, $C_{22}$ and $C_{30}$.}
%\label{fig:OrbitComparFig}
%\end{figure}

\section{Conclusion \& Perspectives}
\label{sec:ConcPersp}

In the frame of this work, we have simulated the BepiColombo rotation experiment and its potential outcome. The BepiColombo orbit characteristics proposed by ESA (\citep{CREMA} (see tab.\ref{tab:allMPOorbitsV})) were used in order to show their impact on the quality of the parameter restitution using the libration experiment observables. By projecting the spacecraft orbit on the surface of the planet and considering surface illumination conditions required for the image matching process, the distribution of potential images on the surface has been obtained.  The uncertainty on the satellite position and orientation (imposed by the precision of the star tracker and radio science observations respectively) is included by subjecting each potential image to a positioning error. In practise, the location and time of possible repeated photographic measurements of selected target positions on the surface of Mercury have been simulated and used as input for the determination of the rotation parameters. Additional errors arise when comparing two images of the same target region on the planetary surface. This process is limited by the photographic resolution of the least well-resolved image and depends on the solar illumination angle of the planetary surface region observed. 

The location and time of possible repeated photographic measurements of selected target positions on the surface of Mercury have been simulated and used as input for the determination of the rotation parameters. By means of a least squares process, three libration amplitudes and the obliquity are estimated.

%In practise, we have simulated repeated photographic measurements of selected target positions on the surface of Mercury in order to estimate 

The accuracy of the reconstruction of the rotational motion of the planet has been studied as a function of the quantity of measurements made, the number of different targets considered, the mission duration, the illumination conditions, the existence of observational altitude limitations and the size of potential spacecraft science blackout zones. Several parameters influence the quality of the libration and obliquity restitution directly and indirectly through the limitation of the surface observability in time and space. Our results can be used to determine the sweet spot in observations and target numbers needed to achieve a good result while keeping the data volume to a minimum (because dedicated imaging opportunities will be limited), and can serve as a measure of the cost of additional observation in contrast to the gains obtained.

As shown in section \ref{sec:SurfaceObs} the nominal image illumination constraints used in this study greatly reduce the surface regions available for target selection as well as the number of image pairs that can be made from these targets over a nominal mission duration of 360 days. This explains the relative insensitivity of the observable surface regions to changes in blackout-zone size, as their consequences manifest themselves mainly outside of the observable regions. In  any case most image pairs are to be found at latitudes between 30$\degr$ North and 30$\degr$ South and in two \emph{windows} centred around the planetary longitude $0\degr$ and $180\degr$, due to the strict illumination conditions. Fortunately as the libration is a longitudinal motion, shifts between images are larger at lower latitudes.  Therefore, images of near-equatorial regions will be of greater importance to the rotational experiment.

The precision on the estimated libration amplitude also depends on the number of possible images of the same target used. More surveyed targets provide a larger data volume and more opportunities for \emph{high-signal} image-pairs, but are limited by the data volume that will be allocatable to the rotation experiment from total BepiColombo mission data volume. 

The criteria for the best determination of the obliquity are different from those of the libration amplitudes because low altitude observations are near symmetrical with respect to Mercury's apohelion and therefore don't yield sufficient information on the obliquity in contrast with the annual libration. The incompatible optimal requirements for obliquity and libration set additional constraints on the strategy to obtain both quantities from this image based experiment, such as the requirement to use high altitude images taken on the MPO's apoherm side in order to be able to solve for the obliquity.

We have shown that, in order to achieve the rotation experiment goals, if a science blackout zone has to be implemented it cannot encompass the planetary perihelion, under penalty of rendering the obliquity estimation impossible, and heavily degrading all libration estimates. Surface observations between $-10\degr$ and $+10\degr$ of Mercury true anomaly are critical to correctly estimate Mercury's obliquity and in turn better constrain the annual and planetary librations.

The two long period planetary induced librations can only be reliably solved for, if the annual libration amplitude \underline{and} the obliquity are well determined.
 
The quality of the spacecraft positioning is very important since it greatly influences the ability to position the images taken by the spacecraft on the planetary surface and hence to estimate the planetary rotation. It is worth to remember that we have considered a worst case scenario for the magnitudes of each individual observation error. This was done in order to guarantee not to be overly optimistic in terms of the quality of the estimation process because of the relatively \emph{basic} error models used.  

%Our simulations show that the achievable level of accuracy on the libration amplitude and obliquity will be sufficient to constrain the size and physical state of the core of Mercury. If the orbiter follows the ESA baseline mission scenario and at least 50 landmarks are imaged at least twice over the mission duration, the libration amplitude can be determined in two Mercury years (176 days) with an accuracy of 3 arcseconds or better.

For the nominal mission scenario we have shown that:
\begin{enumerate}	
	\item The obliquity can be accurately solved for if surface observations around Mercury perihelion can be guaranteed. This is true even for a small number of 25 targets surveyed.
  \item The annual libration amplitude is well solved for in almost all circumstances reaching error levels of around 1 arcsecond in true and formal error after less than 360 days and for data sets limited to 25 targets only.
  \item The main Jupiter and Venus induced libration amplitudes are best estimated for longer mission durations because of their long periods. 
	\item The amplitude of the Jupiter induced libration is expected to be of the order of the annual libration amplitude, because the free libration period of Mercury is expected very close to the Jupiter induced libration period \citep{2008CeMDA.101..141D}. A good estimation of the Jupiter libration amplitude is possible under nominal conditions (as long as the obliquity is well resolved) and could deliver a constraint on the value of the free libration period. If the pointing error can be reduced to its minimum level of $2.5$as the Jupiter amplitude \emph{could} be determined \emph{within} the nominal mission duration ($360$ days) to below $<2.5$as. 
	\item The Venus libration amplitude is ten times smaller than the annual libration amplitude and cannot (even for longer mission durations) be solved for accurately enough to deliver an independent constraint on the interior parameters of Mercury. The error levels never dip below 50$\%$ of the libration amplitude.
	\item All these results can be obtained when considering the worst case camera pointing error level of $5.0$as
\end{enumerate}
For all these reasons we can conclude that the nominal mission scenario as currently defined by ESA, taking into account the possible need for science blackout zones, is able to solve for the obliquity and annual libration amplitude and almost for the Jupiter induced libration amplitude, given that 25 targets are observed repeatedly over the nominal mission of 360 days, and that surface observations close to Mercury perihelion are possible.

\section*{Acknowledgements}
This work was financially supported by the Belgian PRODEX program managed by the European Space Agency in collaboration with the Belgian Federal Science Policy Office.
The authors would like to thank the anonymous reviewers for their constructive comments.

\bibliographystyle{elsarticle-harv}

\bibliography{MaBibT} % my references .bib file

\end{document}